\documentstyle[aps]{revtex}
\input{psfig}

\begin{document}

\baselineskip 10pt

\title{{\large {\bf Cosmic microwave background: 
polarization and temperature anisotropies from symmetric 
structures}}}

\author{Carlo Baccigalupi}
\address{SISSA/ISAS, Via Beirut 4 34014 Trieste, Italy}

\maketitle

\begin{abstract}
Perturbations in the Cosmic Microwave Background (CMB) are 
generated by primordial inhomogeneities. I consider the case 
of CMB anisotropies from one single ordered perturbation source, 
or seed, existing well before decoupling between matter and 
radiation. Such structures could have been left by high energy 
symmetries breaking in the early universe.

I focus on the cases of spherical and cylindrical symmetry of the seed. 
I give general analytic expressions for the polarization and
temperature linear perturbations, 
factoring out of the Fourier integral
the dependence on the photon propagation direction
and on the geometric coordinates describing the seed. 
I show how the CMB perturbations manifestly reflect the symmetries 
of their seeds. In particular, polarization is uniquely linked 
to the shape of the source because of its tensorial nature. 

CMB anisotropies are obtained with a line of sight integration. 
They are function of the position and orientation of the seed 
along the photons path. 

This treatment highlights the undulatory properties 
of the CMB. I show with numerical examples how the
polarization and temperature perturbations 
propagate beyond the size of their 
seeds, reaching the CMB sound horizon
at the time considered. Just like the waves from a pebble
thrown in a pond, CMB anisotropy from a seed intersecting the 
last scattering surface appears as a series of temperature and 
polarization waves surrounding the seed, 
extending on the scale of the CMB sound horizon at decoupling, 
roughly $1^{o}$ in the sky. Each wave is characterized by its own 
value of the CMB perturbation, with the same mean amplitude 
of the signal coming from the seed interior; 
as expected for a linear structure with size $L\le H^{-1}$ 
and density contrast $\delta$ at decoupling, the 
temperature anisotropy is $\delta T/T\simeq \delta (L/H^{-1})^{2}$, 
roughly ten times stronger than the polarization. 

These waves could allow to distinguish 
relics from high energy processes of 
the early universe from point-like 
astrophysical sources, because of their angular extension 
and amplitude. Also, the marked analogy between polarization and
temperature signals offers cross correlation possibilities 
for the future detection instruments. 
It would be interesting to detect these signals in the 
next $10'$ CMB map provided by the Planck 
Surveyor satellite experiment. 


\noindent
FERMILAB-Pub-98/204-A
\end{abstract}

\section{Introduction}
\label{introduction}
The cosmic microwave background (CMB) carries 
detailed information about the high energy physical processes
occurred in the early universe. Most probably, the microphysics 
still hidden to our knowledge left traces that have been stretched 
out to large and observable scales by a period of accelerated expansion; 
at decoupling between matter and radiation, they imprinted
anisotropies in the CMB. This is the reason of the contemporary 
theoretical and experimental efforts to understand the CMB physics. 
The theory of the CMB anisotropies has been deeply explored 
in the past (see \cite{BKS,MB} and references therein) and, 
recently, it has been casted in a complete and organic 
form \cite{HW}. At the same time, 
many experiments are at work to explore the CMB anisotropies 
toward smaller and smaller angular scales (see \cite{CMBPAST} 
for reviews); this experimental enterprise will culminate with 
the Planck mission of the next decade, that will provide
the whole sky temperature and polarization anisotropy map 
down to a minimum detectable perturbation of one part over 1 million 
and with an angular resolution of about $10'$ \cite{PLANCK}.

According to the inflationary phenomenology, a scalar
field (the inflaton) slowly rolls toward the minimum of 
its potential, giving the non-zero vacuum energy responsible 
for the expansion itself. 
The quantum fluctuations are thought to arise from the vacuum 
in a curved background; they are 
stretched out to large scales by the inflationary expansion 
itself, and set up the 
seeds of the cosmological perturbations we observe today 
(see \cite{MFB} for reviews). However, even adopting this 
inflationary scenario, things are still unclear 
for what concerns the release of the 
energy stored in the inflaton into ordinary matter and 
radiation, the so called reheating (or {\it pre}heating) era 
\cite{PRE}. The oscillations of the inflaton around 
its minimum, combined with the coupling to other fields, 
can restore high energy symmetries that have to be broken to 
reach our low energy minimum; consequently, 
a post-inflationary generation of topological defects may 
arise, and this occurrence is at the present under investigation 
\cite{TD}. Also, during inflation itself many fundamental 
fields may act on stage and the effective potential may have 
several minima separated by potential barriers. 
If this is the case, tunneling
phenomena occur, and the nucleated bubbles are stretched out 
to large scales as the ordinary quantum fluctuations 
(see \cite{FOI} for reviews); 
at reheating the energy stored in the 
shells is converted into matter and 
radiation and bubbly traces may be left 
in the density distribution 
(this possibility, with different points of view, 
has been considered in the last decade \cite{BUBBLE}). 

Suppose that one of these relics from very high energy physics 
is plunged from some very early time into cosmic matter and radiation, 
no matter of its composition, that could be scalar field or 
cosmic fluid or other. 
It generates perturbations around itself, in particular  
in the photon-baryon fluid. If also it intersects the last 
scattering surface 
(LSS, the place of origin of the CMB), 
these perturbations become anisotropies that we could 
observe today. These are expected to be well recognizable, 
since in most 
cases such seed is a spatially limited structure, very
different from the diffuse fluctuations of the
pure slow-roll inflation; technically speaking, such signal
would be strongly non-Gaussian and non-scale-invariant. 
Also, such structures are expected to possess (approximate) symmetries, 
like a bubble or a monopole (spherical) and a string (cylindrical). 
Their detection in the CMB anisotropies would be 
the first observational evidence of the existence of 
high energy symmetries, and this hope is precisely
the motive of this work. I develop here some useful 
formulas for the CMB perturbations and anisotropies from 
symmetric structures; the results are {\it independent} 
from the particular seed, the 
only characterization being its 
symmetry, that I take here spherical 
or cylindrical. I perform some 
numerical integrations using these 
formulas and adopting toy symmetric seeds, in order to 
investigate the geometrical and dynamical properties of their 
own CMB perturbations and anisotropies. In forthcoming works 
I will compute the CMB anisotropies 
from realistic relics left from 
high energy physics in the early universe; 
a pretty example, valid simply for large 
bubbles in the density distribution, may be found in \cite{B}. 

As already mentioned, the treatment of the CMB inhomogeneities has 
been casted recently in a complete and organic form, the
total angular momentum method \cite{HW}. In turn, 
it is based on the general treatment of the linear cosmological 
perturbations \cite{BKS}; I perform the calculations in this
frame, respecting the notations as much as possible. 
The CMB perturbations involve temperature 
($\delta T/T\equiv \Theta$ in the following) and 
polarization, that is expressed via the Stokes parameters
$Q$ and $U$ describing linear polarization. 
For a given Fourier mode specified by the $\vec{k}$ vector, 
it is convenient to express the relevant quantities
in a frame in which the $\hat{k}$ direction is the polar axis
(the $\hat{k}$-frame in the following). The reason is that, 
in the new frame, the scalar, vector and tensor components of 
the perturbed metric quantities are coupled respectively 
to the $m=0,\pm 1,\pm 2$ indexes of the spherical 
harmonics \cite{HW}. Of course, transforming 
back to the real space, the $\hat{k}$-frame quantities must be
expressed in the fixed laboratory frame (the $lab$-frame in the
following). For a given Fourier mode $\vec{k}$, 
$Q$ is the difference in temperature fluctuations polarized
in the $\hat{e}_{\theta}$ and $\hat{e}_{\phi}$ directions 
($\theta$ and $\phi$ being the usual angles in spherical coordinates); 
$U$ is the same difference where the axes have been rotated 
by 45$^{o}$ around the photon propagation direction.
Equivalently, $Q$ and $U$ may be seen as the expansion 
coefficients of the polarization tensor into the Pauli 
matrices $\sigma_{3}$ and $\sigma_{1}$, defined on the 
basis vectors $\hat{e}_{\theta}$ and $\hat{e}_{\phi}$ 
in the $\hat{k}$-frame.

The background Friedmann Robertson Walker (FRW) 
metric is
\begin{equation}
\label{frw}
ds^{2}=a(\eta )^{2}\left(-d\eta^{2}+
{dr^{2}\over 1-Kr^{2}}+r^{2}d\Omega^{2}\right)\ ,
\end{equation}
where $\eta (t)=\int_{0}^{t} d\tau/a(\tau )$ is the 
conformal time and $K$ the spatial curvature; 
I will assume a flat $K=0$ background in this work. 
The perturbed metric tensor is
\begin{equation}
\label{frwpert}
g_{\mu\nu}=a(\eta )^{2}(\gamma_{\mu\nu}+h_{\mu\nu})\ ,
\end{equation}
where $a(\eta )^{2}\gamma_{\mu\nu}$ represents the background. 
Since $h_{\mu\nu}\ll\gamma_{\mu\nu}$, a gauge freedom reduces 
the number of physically significant quantities in 
the perturbation metric tensor; 
in this work I adopt the generalized Newtonian gauge in which 
the two scalar perturbed metric component are $\Psi =h_{00}/2$ 
and $2\Phi =h_{11}=h_{22}=h_{33}$ \cite{BKS,HW}. 

The CMB perturbations depend 
on the spacetime point and on the photon propagation direction 
$\hat{n}$, so an appropriate normal mode expansion is needed:
\begin{equation}
\label{dttdeveloped}
\Theta (\eta ,\vec{r},\hat{n}) = \int{d^{3}k\over (2\pi)^{3}}
\sum_{l}\sum_{m=-2}^{2}
\Theta_{l}^{(m)}(\eta ,\vec{k})G_{l}^{m}(\vec{r},\vec{k},\hat{n})\ ,
\end{equation}
$$
(Q+iU)(\eta ,\vec{r},\hat{n}) {\bf M}_{+} +
(Q-iU)(\eta ,\vec{r},\hat{n}) {\bf M}_{-} =
$$
\begin{equation}
\label{qudeveloped}
\int{d^{3}k\over (2\pi)^{3}}
\sum_{l\ge 2}\sum_{m=-2}^{2}
\left[(E_{l}^{(m)}+iB_{l}^{(m)})(\eta ,\vec{k})_{+2}G_{l}^{m}
(\vec{r},\vec{k},\hat{n})
+(E_{l}^{(m)}-iB_{l}^{(m)})(\eta ,\vec{k})_{-2}G_{l}^{m}
(\vec{r},\vec{k},\hat{n})\right]\ ,
\end{equation}
where ${\bf M}_{\pm}=(\sigma_{3}\mp i\sigma_{1})/2$ are
convenient basis matrices for the polarization tensor. 
$G_{l}^{m}$ and $_{\pm 2}G_{l}^{m}$ include both spatial 
and angular functions; the spatial ones are the eigenmodes 
of the Laplacian in the metric (\ref{frw}):
\begin{equation}
\label{lap}
\nabla^{2}Q_{K}(\vec{k},\vec{x})\equiv
\gamma^{ij}Q_{K\ |ij}=-k^{2}Q_{K}(\vec{k},\vec{x})\ ;
\end{equation}
the angular functions are instead spherical harmonics.
In the case of flatness ($K=0$) the Laplace equation (\ref{lap})
gives plane waves, and the expression of the normal modes becomes
\begin{equation}
\label{glms}
G_{l}^{m}=(-i)^{l}\sqrt{4\pi\over 2l+1}
Y_{l}^{m}(\hat{n}_{\hat{k}})\exp(i\vec{k}\cdot\vec{r})\ ,
\end{equation}
\begin{equation}
\label{glmt}
_{\pm 2}G_{l}^{m}=(-i)^{l}\sqrt{4\pi\over 2l+1}
\, _{\pm 2}Y_{l}^{m}(\hat{n}_{\hat{k}}){\bf M}_{\pm}^{\hat{k}}
\exp(i\vec{k}\cdot\vec{r})\ ;
\end{equation}
as a difference with respect to \cite{HW}, 
the notation $\hat{n}_{\hat{k}}$, ${\bf M}_{\pm}^{\hat{k}}$ 
has been used to underline that, for each $\vec{k}$ mode, all 
the quantities in (\ref{glms},\ref{glmt}), 
as well as the expansion coefficients in 
(\ref{dttdeveloped},\ref{qudeveloped}), are expressed in the 
$\hat{k}$-frame; as customary, the 
expansion coefficients of the Stokes parameters $Q,U$ 
have been decomposed into real and imaginary parts. 
Throughout this work, in order to characterize the polarization 
within symmetric seeds, I make use of the 
useful definition of {\it polarization 
direction} \cite{BW}, given entirely in terms of $Q$ and $U$ 
as follows. It is easy to see that, due to the rotation 
properties of the Pauli matrices, the angle 
\begin{equation}
\label{pd}
\alpha ={1\over 2}{\rm tan}^{-1}{U\over Q}
\end{equation}
goes into $\alpha -\phi$ under a rotation by $\phi$ around 
$\hat{n}$; thus it defines a fixed axis on the plane orthogonal 
$\hat{n}$, that is the polarization direction. 

The underlying cosmological inhomogeneities move the CMB 
perturbations and are encoded in the 
expansion coefficients in (\ref{dttdeveloped},\ref{qudeveloped}). 
Before going to the content of this work, it is useful 
to point out the following important distinction. 
The Fourier transform of any perturbation 
quantity $\Delta$ may be written as
\begin{equation}
\label{ft}
\Delta (\vec{k})=|\Delta (\vec{k})|e^{i\phi_{\vec{k}}}\ ;
\end{equation}
it is Gaussian if the phases in $\phi_{\vec{k}}$ are random; 
specifically in these hypothesis, the statistics is completely 
described by the power spectrum, $<|\Delta (\vec{k})|^{2}>$. 
Also it is scale-invariant if the modulus depends only on the scale 
($k=|\vec{k}|$) in such a way that the power associated to each one 
is the same at the horizon reenter. 
On the contrary, CMB anisotropies from sources like
the ones considered here are non-Gaussian and non-scale-invariant; 
their symmetries, encoded in precise 
properties of both modulus and phases in (\ref{ft}), 
are their unique sign in the CMB. Moreover, I do not 
require that they are dominant for structure formation. 
An high resolution CMB map could contain the unambiguous 
imprint of one single symmetric seed existing at decoupling 
plunged in a global Gaussian signal; even if the power spectrum 
does not contain its sign at all, that would be enormously 
interesting! 

The work is organized as follows.
Sections II and III contain the analysis of the CMB perturbations 
in spherical and cylindrical symmetry respectively.
Section IV contains the method for the 
computation of the CMB polarization and 
temperature anisotropies as they would appear on the sky. 
In section V the results from numerical integrations are shown. 
Finally, section VI contains the conclusions. 

\section{Spherical symmetry}

It is easy to see that spherical structures may be scalar only, 
and thus are described by the $m=0$ modes of the linear expansion;
there is no way to comb the hair of a sphere in such
a way to obtain a spherical distribution, and this prevents 
spherical structures to be made of genuinely vector 
(or tensor) components. Thus I drop the $^{(0)}$ index in the 
following, and consider flat space geometry, $K=0$.

The problem to solve is the following: at a conformal time 
$\eta$, a perfect CMB detector is placed 
in a point $\vec{r}$ nearby a primordial spherical structure; 
what's the CMB perturbations carried by photons scattered on a 
direction $\hat{n}$? 

The center of the coordinate frame is placed at the center of 
the spherical seed. Its Fourier transform depends only on the 
wavevector modulus $k$ and it is therefore the same 
for any axes orientation: 
\begin{equation}
\label{sphere}
\vec{r}\rightarrow r\ \ \Leftrightarrow\ \ \vec{k}\rightarrow k\ .
\end{equation}
First, let us find the consequence of (\ref{sphere}) on the 
pure temperature perturbation $\Theta$. 
The expansion coefficients $\Theta_{l}$ in (\ref{dttdeveloped})
are proportional to the Fourier transformed perturbation 
(see section V) and do not depend at all on the orientation of the 
perturbation in the $\hat{k}$-frame, simply because it is spherical:
they depend on $k$ only. Consequently, posing $d^{3}k=k^{2}dk
d\Omega_{\hat{k}}$ in (\ref{dttdeveloped}) the $\Theta_{l}$ 
coefficients may be extracted from the angular integral. 
Thus let us face the pure geometric quantity
\begin{equation}
\label{int1}
\int d\Omega_{\hat{k}}
(-i)^{l}\sqrt{4\pi\over 2l+1}Y_{l}^{0}(\hat{n}\cdot\hat{k})
\exp(i\vec{k}\cdot\vec{r})\ ,
\end{equation}
where the argument of the spherical harmonics in 
(\ref{glms}) has been shown (see appendix A). 
The integral (\ref{int1}) is easily computed expanding 
the plane wave into Bessel and Legendre functions
\begin{equation}
\label{wavebessel}
e^{i\vec{k}\cdot\vec{r}}=
\sum_{l}
i^{l}(2l+1)j_{l}(kr)P_{l}(\hat{r}\cdot\hat{k})\ ,
\end{equation}
and employing the useful relation (\ref{legendre})
with $\hat{n}_{1}=\hat{k}$, $\hat{n}_{2}=\hat{r}$ and 
$\hat{n}_{3}=\hat{n}$. The result is:
\begin{equation}
\label{thetasphere}
\Theta =
\sum_{l}
P_{l}(\hat{n}\cdot\hat{r})
\int {k^{2}dk\over 2\pi^{2}}\Theta_{l}(\eta ,k)j_{l}(kr)\ .
\end{equation}
This expression gives the CMB temperature perturbation 
at any time for the most general spherical 
perturbation, encoded in the Fourier integral. 
The dependence on $\hat{n}$ and $\hat{r}$ 
has been factored out, and enters only in the Legendre polynomials
argument $\hat{n}\cdot\hat{r}$. This is an expected feature 
of this spherical case: for example, focus on the $l=1$ term, 
better known as the Doppler effect ($\Theta_{1}$ is essentially the 
velocity of baryons \cite{MB}); the motion of each 
particle in this spherical case is radial of course; then, 
since this Legendre polynomial is just $\hat{n}\cdot\hat{r}$, 
photons propagating on the direction $\hat{n}$ pick up 
the usual Doppler cosine contribution at the scattering point. 

Let us face now the polarization for 
a spherical seed. A first simplification is that the scalar 
perturbations excite 
the $E_{l}$ modes only \cite{HW}, so we can drop the $B_{l}$ terms in 
the following. Then, as before, the $E_{l}$ coefficients depend
on $k$ only, so they can be extracted from the angular integral. 
As a difference from the temperature case, the {\it tensor} spherical 
harmonics describe now the angular dependence in (\ref{glmt}); 
fortunately they admit, for $m=0$, a simple expression in terms of the 
elementary Legendre polynomials, as it is demonstrated in appendix A: 
\begin{equation}
\label{tensorlegendre}
_{2}Y_{l}^{0}(\hat{n}\cdot{\hat{k}})=
\sqrt{{2l+1\over 4\pi}{(l-2)!\over (l+2)!}}
P_{l}^{2}(\hat{n}\cdot\hat{k})\ ,
\end{equation}
where the $\pm$ index has been suppressed since it makes no 
difference in the $m=0$ case. 

Focus now on the ${\bf M}_{\pm}^{\hat{k}}$ matrices.
They have to be expressed in terms of the fixed $lab$-frame matrices 
${\bf M}_{\pm}$. This is obtained performing a 
rotation around the $\hat{n}$ axis in order to make the 
$\hat{e}_{\theta}$ and the $\hat{e}_{\phi}$ 
vectors in the $\hat{k}$-frame coincident with the 
laboratory ones: the rotation angle is essentially the angular 
coordinate of the projection of $\hat{k}$ 
into the plane orthogonal to $\hat{n}$. 
For simplicity, but without any loss of generality, 
let's orient the $lab$-frame so that $\hat{n}$ is the 
polar axis; then, it is easy to see that the rotation angle is simply 
$-(\phi_{\hat{k}}+\pi )$, where $\phi_{\hat{k}}$ is just the $\phi$ 
coordinate of $\vec{k}$ in the $lab-$frame; 
thus, from elementary rotation properties 
of the Pauli matrices, the expression of ${\bf M}_{\pm}^{\hat{k}}$ 
as seen in the $lab-$frame is 
\begin{equation}
\label{kpolarlab}
{\bf M}_{\pm}^{\hat{k}}=
e^{\mp 2i\phi_{\hat{k}}}{\bf M}_{\pm}\ .
\end{equation}
The integral in (\ref{qudeveloped}) has now the following form:
\begin{equation}
(Q\pm iU){\bf M}_{\pm}=
\label{intpol1}
\sum_{l\ge 2}
\int{k^{2}dk\over (2\pi )^{3}}
E_{l}(\eta ,k)(-i)^{l}
\sqrt{(l-2)!\over (l+2)!}\int d\Omega_{\hat{k}}
e^{i\vec{k}\cdot\vec{r}}
P_{l}^{2}(\hat{n}\cdot\hat{k})\cdot
e^{\mp 2i\phi_{\hat{k}}}{\bf M}_{\pm}\ .
\end{equation} 
Moreover, it is useful to employ the expansion (\ref{wavebessel}) 
together with the addition relation (\ref{addition}) with 
$\hat{n}_{1}=\hat{r}$ and $\hat{n}_{2}=\hat{k}$:
\begin{equation}
\label{auxi}
e^{i\vec{k}\cdot\vec{r}}=
\sum_{l,m=-l}^{l}
i^{l}4\pi j_{l}(kr)Y_{l}^{m}(\hat{r})Y_{l}^{m*}(\hat{k})\ .
\end{equation}
The integral (\ref{intpol1}) in $d\phi_{\hat{k}}$ can now be calculated: 
the $e^{\mp 2i\phi_{\hat{k}}}$ phases in (\ref{kpolarlab}) 
select the $m=\mp 2$ terms respectively above; 
once this is done, the integral on $d\theta_{\hat{k}}$ 
is simple using the spherical harmonics orthogonality (\ref{orthoplm}); 
the final result is 
$$
(Q+iU){\bf M}_{+}+(Q-iU){\bf M}_{-}=
$$
\begin{equation}
\label{intpol}
\left(e^{-2i\phi_{\hat{r}}}{\bf M}_{+}+
e^{2i\phi_{\hat{r}}}{\bf M}_{-}\right)\cdot
\sum_{l\ge 2}
\sqrt{(l-2)!\over (l+2)!}
P_{l}^{2}(\hat{n}\cdot\hat{r}) 
\cdot\int{k^{2}dk\over 2\pi^{2}}
E_{l}(\eta ,k)j_{l}(kr)\ ,
\end{equation} 
where $\phi_{\hat{r}}$ is the angular coordinate 
of the projection of the $\vec{r}$ vector on the plane orthogonal 
to $\hat{n}$. Let's check out the meanings of (\ref{intpol}). 
Again the dependence on $\hat{n}$ and $\hat{r}$ has been completely 
extracted from the Fourier integral; really the matrices ${\bf M}_{\pm}$, 
basis for the polarization tensor, are outside the sum on $l$ and 
multiply appropriate phases: this makes 
easy the following geometric consideration. 
If we choose the $lab$-axes so that $\phi_{\hat{r}}=0$, 
the matrix in (\ref{intpol}) is simply 
${\bf M}_{+}+{\bf M}_{-}=\sigma_{3}$ and the polarization 
quantities results in a pure $Q$ term; 
thus (\ref{intpol}) gives the difference in the polarization
amplitudes relative to the axes displayed in the upper
panel of figure 1, one lying on the plane formed by the 
$\hat{n}$ and $\hat{r}$ directions and the other orthogonal 
to the same plane. With this axes orientation, the angle $\alpha$ in 
(\ref{pd}) is zero: this means that the polarization direction within 
a spherical seed lies on the plane formed by $\hat{n}$ and $\hat{r}$, 
as sketched in figure 1. 
As a related important point, note the second order Legendre 
polynomial $P_{l}^{2}$ (the temperature case had $P_{l}$); it is 
meaningful since it guarantees that light propagating radially
is not polarized ($P_{l}^{2}\propto \sin^{2}{\theta}P_{l}$):
the radial propagation in spherical symmetry is an
axial symmetric problem, so that no preferred direction exists for
the polarization, since it belongs on the plane orthogonal to the
symmetry axis. 

These results, together with the 
temperature ones, completely characterize 
the CMB perturbation carried by 
photons moving in a spherical seed, 
independently from any other specification. 
The next section contains the same 
analysis developed here, but based on cylindrical seeds. 

\section{Cylindrical symmetry}

Scalars can be arranged cylindrically of course, 
but also vectors (vorticity is a vectorial feature). 
Consequently, the $m=0,\pm 1$ are allowed. In the vector 
case however, the generic $\vec{k}$ mode of the Fourier 
transform is a vector of course; thus its orientation enters 
in the angular integrals of (\ref{dttdeveloped}) and 
(\ref{qudeveloped}), that become strongly dependent on the 
particular seed considered. For this reason, again 
I restrict to the scalar case (dropping the $^{(0)}$ index), 
and employ flat FRW, $K=0$. 

The Fourier transform of a cylindrically symmetric quantity 
$\Delta$ may be expressed as 
\begin{equation}
\label{cylft}
\Delta (\eta ,\vec{k})\equiv
\Delta (\eta ,k_{+},k_{z})\ ,
\end{equation}
where $k_{z},k_{+}$ are the component of 
$\vec{k}$ on the symmetry axis and on the equatorial 
plane respectively, $k_{+}^{2}=k_{x}^{2}+k_{y}^{2}$. 
If the seed is invariant under traslations along the symmetry 
axis (this case is mentioned as infinite cylindrical seed in 
the following), then the expression is 
\begin{equation}
\label{cylftl}
\Delta (\eta ,\vec{k})\equiv
\Delta (\eta ,k_{+})\cdot 2\pi\delta(k_{z})\ ,
\end{equation}
where $\delta (k_{z})$ is the Dirac delta.
Consequently, the expansion coefficients of 
(\ref{dttdeveloped}) and (\ref{qudeveloped}) depend
only on $k_{+},k_{z}$ in the first case and on 
$k_{+}$ in the second. 

First, consider the temperature perturbation. 
In cylindrical coordinates 
$d^{3}k=k_{+}dk_{+}dk_{z}d\phi_{\hat{k}}$ 
and the $\Theta_{l}$ coefficients come 
out of the integral in $d\phi_{\hat{k}}$:
\begin{equation}
\label{cylint}
\Theta =\sum_{l}
\int{k_{+}dk_{+}dk_{z}\over (2\pi)^{3}}
\Theta_{l}(\eta ,k_{+},k_{z})\int d\phi_{\hat{k}}
(-i)^{l}\sqrt{4\pi\over 2l+1}Y_{l}^{0}
(\hat{k}\cdot\hat{n})e^{i\vec{k}\cdot\vec{r}}\ .
\end{equation}
Again I make use of the addition relation 
(\ref{addition}), choosing the polar axis 
in the $lab$-frame 
coincident with the symmetry axis; 
I apply it twice, both on the $Y_{l}^{0}=
\sqrt{4\pi /(2l+1)}P_{l}^{0}$ 
above and on the plane wave, 
expanded as (\ref{auxi}). Paying the 
price to increase the number of sums, 
equation (\ref{cylint}) becomes
$$
\Theta =\sum_{l,m=-l}^{l}
(-i)^{l}{4\pi\over 2l+1}Y_{l}^{m}(\hat{n})\cdot
\sum_{l',m'=-l'}^{l'}
i^{l'}4\pi Y_{l'}^{m'}(\hat{r})\cdot
$$
\begin{equation}
\label{cylgeo}
\cdot\int {k_{+}dk_{+}dk_{z}\over (2\pi)^{3}}
\Theta_{l}(\eta ,k_{+},k_{z})j_{l'}(kr)
\int d\phi_{\hat{k}}
Y_{l}^{m*}(\hat{k})Y_{l'}^{m'*}(\hat{k})\ ,
\end{equation}
where of course $k^2=k_{+}^{2}+k_{z}^{2}$ 
and $r^{2}=r_{+}^{2}+r_{z}^{2}$.
Now, it is manifest that the 
integral in $d\phi_{\hat{k}}$ kills
everything except for the $m'=-m$ 
terms; thus, the phase of 
$Y_{l'}^{m'}(\hat{r})$ precisely 
fits together with the 
phase of $Y_{l}^{m}(\hat{n})$, 
making them relative. 
Writing in full the spherical harmonics, the final result is: 
$$
\Theta =\sum_{l,m=-l}^{l}
(-i)^{l}P_{l}^{m}(\hat{n}\cdot\hat{z})
e^{im(\phi_{\hat{n}}-\phi_{\hat{r}})}
\sum_{l'\ge |m|}
i^{l'}(2l'+1) P_{l'}^{-m}(\hat{r}\cdot\hat{z})\cdot
$$
\begin{equation}
\label{cylt}
\cdot\int{k_{+}dk_{+}dk_{z}\over 4\pi^{2}}
\Theta_{l}(\eta ,k_{+},k_{z})j_{l'}(kr)
P_{l}^{-m}(\hat{k}\cdot\hat{z})P_{l'}^{m}(\hat{k}\cdot\hat{z})\ . 
\end{equation}
It is useful to note that writing $\vec{k}\cdot\vec{r}=
\vec{k}_{+}\cdot\vec{r}_{+}+k_{z}z$ and extracting the exponential 
regarding the last term  
from the integral in $d\phi_{\hat{k}}$, the expression 
(\ref{cylt}) may be simplified:
$$
\Theta =\sum_{l,m=-l}^{l}
(-i)^{l}P_{l}^{m}(\hat{n}\cdot\hat{z})
e^{im(\phi_{\hat{n}}-\phi_{\hat{r}})}
\sum_{l'\ge |m|}
i^{l'}(2l'+1)P_{l'}^{m}(0)P_{l'}^{-m}(0)\cdot
$$
\begin{equation}
\label{cyltsimple}
\cdot\int{k_{+}dk_{+}dk_{z}\over 4\pi^{2}}
\Theta_{l}(\eta ,k_{+},k_{z})e^{ik_{z}z}j_{l'}(k_{+}r_{+})
P_{l}^{-m}(\hat{k}\cdot\hat{z})\ ; 
\end{equation}
note that the sum over $l'$ is restricted to the even $l'+m$ 
terms because of equation (\ref{legendre0m}).

As an alternative approach, one can give up the expansion
of the plane wave in (\ref{cylint}), and express its argument 
as $\vec{k}\cdot\vec{r}=k_{+}r_{+}
\cos (\phi_{\hat{k}}-\phi_{\hat{r}})+k_{z}z$. 
As above, the phase of $Y_{l}^{m*}(\hat{k})$ is 
$e^{-im\phi_{\hat{k}}}=e^{im(\phi_{\hat{r}}-\phi_{\hat{k}})}
\cdot e^{-im\phi_{\hat{r}}}$; the second factor comes out of
the integral, and again fits together with the 
corresponding phase of $Y_{l}^{m}(\hat{n})$. 
The advantage of this approach is that the integral in 
$d\phi_{\hat{k}}$ has a note form and the result is: 
\begin{equation}
\label{cyltaw}
\Theta =\sum_{l,m=-l}^{l}(-i)^{l}
P_{l}^{m}(\hat{n}\cdot\hat{z})\cdot 
e^{im(\phi_{\hat{n}}-\phi_{\hat{r}})}
\int{k_{+}dk_{+}dk_{z}\over (2\pi)^{3}}
e^{ik_{z}z}\Theta_{l}(\eta ,k_{+},k_{z})
P_{l}^{-m}(\hat{k}\cdot\hat{z}){\bf JE}_{+}(m,k_{+}r_{+})\ ,
\end{equation} 
where the function ${\bf JE}_{+}$ is a combination of the
Anger and Weber functions, defined in appendix B.

The expressions corresponding to (\ref{cylt},\ref{cyltsimple}) and 
(\ref{cyltaw}) for an infinite cylindrical structure are simpler 
because of the effect of the Dirac delta; 
it eliminates the dependence on $z$, 
reducing the argument of the exponential in (\ref{cylint}) to 
$i\vec{k}_{+}\cdot\vec{r}_{+}$. Also the Legendre polynomials 
into the integral in $d\phi_{\hat{k}}$ have now to be calculated 
for $\hat{k}\cdot\hat{z}=0$, their values being found using
(\ref{legendre0}); from (\ref{legendre0m}) only the terms with 
even $l+m$ and $l'+m$ (and therefore $l+l'$) survive. 
Thus, the expression for $\Theta$ in this case is similar to 
(\ref{cyltsimple}), but simpler:
$$
\Theta =\sum_{l,m=-l}^{l}
P_{l}^{m}(\hat{n}\cdot\hat{z})P_{l}^{-m}(0)
e^{im(\phi_{\hat{n}}-\phi_{\hat{r}})}
\sum_{l'\ge |m|}
(-1)^{l+(l+l')/2}(2l'+1)
\cdot P_{l'}^{m}(0)P_{l'}^{-m}(0)\cdot
$$
\begin{equation}
\label{cylit}
\cdot\int{k_{+}dk_{+}\over 2\pi}
\Theta_{l}(\eta ,k_{+})j_{l'}(k_{+}r_{+})
\ \ \ \ ({\rm even}\ l+m,l'+m,l+l')\ .
\end{equation}
In the second approach
\begin{equation}
\label{cylitaw}
\Theta =\sum_{l,m=-l}^{l}
(-i)^{l}
P_{l}^{m}(\hat{n}\cdot\hat{z})P_{l}^{-m}(0)\cdot 
e^{im(\phi_{\hat{n}}-\phi_{\hat{r}})}
\int{k_{+}dk_{+}\over 4\pi^{2}}
\Theta_{l}(\eta ,k_{+}){\bf JE}_{+}(m,k_{+}r_{+})\ .
\end{equation} 
Let us check the geometric meanings of the above 
expressions. First note how the cylindrical symmetry
caused complications, both in the geometric and integral 
quantities, with respect to the spherical case. However, again 
the dependence on $\hat{n}$ and $\hat{r}$ has
been separated and factored out. The symmetry forces 
the phases of the harmonics with argument $\hat{n}$ 
and $\vec{r}$ to be relative: 
for $r_{+}\ne 0$, the perturbation depends, 
together with the angle between
the symmetry axis and $\hat{n}$, on the direction of 
the projection of $\hat{n}$ on the equatorial plane with 
respect to $\hat{r}_{+}$, as it is intuitive in a 
cylindrical problem; the pure Doppler contribution from the 
peculiar velocity of photons and baryons ($\Theta_{1}$) 
may be easily recognized in the $l=1,m=1$ terms. 
If $\vec{r}$ lies on the symmetry axis itself 
the ${\bf JE}_{+}$ function in (\ref{cyltaw}) 
and (\ref{cylitaw}) reduces simply to $2\pi\delta_{m0}$, 
as shown in appendix B. As a final intuitive feature, 
note how in the case $\hat{n}||\hat{z}$, the
CMB perturbation for an infinitely long seed possesses
a parity symmetry, $\hat{n}\rightarrow -\hat{n}$, since 
all the $m\ne 0$ terms vanish, making $l$ and $l'$ even. 

Let us face now the CMB polarization from cylindrical sources. 
As in the previous section, the $E_{l}$ coefficients come 
out of the integral in $d\phi_{\hat{k}}$ and the tensor harmonics 
are expressed as in (\ref{tensorlegendre}). 
Then the polarization matrices in the $\hat{k}$-frame have to be 
expressed in terms of the corresponding ones in the $lab$-frame: 
${\bf M}_{\pm}^{\hat{k}}=e^{\mp 2i\alpha_{\hat{k}}}{\bf M}_{\pm}$;
as before $\alpha_{\hat{k}}$ is 
the angular coordinate of the projection 
of the $\hat{k}$ versor on the plane orthogonal to $\hat{n}$. 
It is indicated differently 
from (\ref{kpolarlab}) because of the following 
reason. In the previous section 
we were dealing with spherical perturbations;
no matter of how the $lab$-frame 
axes were oriented. This freedom 
allowed us to orient the polar axis 
as $\hat{n}$, so that $\alpha_{\hat{k}}$
was simply related to the $\phi$ coordinate of $\hat{k}$. 
Now things are different: the perturbation source has 
a preferred axis, and the equatorial plane 
is therefore different from the
polarization plane (orthogonal to $\hat{n}$); consequently, 
$\alpha_{\hat{k}}$ depends on $\phi_{\hat{k}}$ in a less simple 
way, as I write below, and this complicates the computations 
of course.

Highlighting again
the integral in $d\phi_{\hat{k}}$, the quantities in 
(\ref{qudeveloped}) take the form
\begin{equation}
\label{cylqudeveloped}
(Q\pm iU){\bf M}_{\pm}=\sum_{l\ge 2}
\int{k_{+}dk_{+}dk_{z}\over (2\pi)^{3}}
E_{l}(\eta ,k_{+},k_{z})
\int d\phi_{\hat{k}}
(-i)^{l}\sqrt{(l-2)!\over (l+2)!}P_{l}^{2}(\hat{n}\cdot\hat{k})
e^{i(\vec{k}\cdot\vec{r}\mp 2\alpha_{\hat{k}})}{\bf M}_{\pm}\ .
\end{equation}
In spite of its innocent appearance, the integral in 
$d\phi_{\hat{k}}$ is not so available 
for extracting the dependence on 
$\hat{n}$ as in the previous cases. 
This is due to the expression of $\alpha_{\hat{k}}$; 
according to the definition 
above, and taking as reference axis the intersection between the 
planes orthogonal to $\hat{n}$ and $\hat{z}$, its expression is
\begin{equation}
\label{cumber}
\cos\alpha_{\hat{k}}={\hat{n}\times\hat{z}\cdot\hat{k}\over
|\hat{n}\times\hat{z}|\sqrt{1-(\hat{k}\cdot\hat{n})^{2}}}\ .
\end{equation}
Thus, $\alpha_{\hat{k}}$ is related to $\phi_{\hat{k}}$ by the 
following relation, that may be easily verified:
\begin{equation}
\label{akpk}
\cos\alpha_{\hat{k}}=\cos\phi_{\hat{k}}
\sqrt{1-(\hat{k}\cdot\hat{z})^{2}\over 
1-(\hat{k}\cdot\hat{n})^{2}}\ .
\end{equation}
Unfortunately, (in my knowledge) there is no simple treatment 
of the angular integral in (\ref{cylqudeveloped}) 
with $\alpha_{\hat{k}}$ given by (\ref{akpk}). 
However, there are some interesting and useful 
particular cases in which 
computations are simpler. First, suppose 
that the photon propagation
direction $\hat{n}$ is parallel to the symmetry axis. Thus 
$\alpha_{\hat{k}}=\phi_{\hat{k}}$, (\ref{kpolarlab}) holds, 
and the Legendre polynomials can be 
extracted from the integral in $d\phi_{\hat{k}}$, 
since now $\hat{n}=\hat{z}$. In the first approach all the 
task consists in expanding the exponential in (\ref{cylqudeveloped}), 
while the second is straightforward. 
The integral precisely kills everything except for the $m=\mp 2$ terms: 
$$
(Q\pm iU){\bf M}_{\pm}=
{\bf M}_{\pm}e^{\mp 2i\phi_{\hat{r}}}\cdot
\sum_{l\ge 2}
(-i)^{l}\sqrt{(l-2)!\over (l+2)!}\cdot
\sum_{l'\ge 2}
i^{l'}(2l'+1)P_{l'}^{\mp 2}(\hat{r}\cdot\hat{z})\cdot
$$
$$
\cdot
\int{k_{+}dk_{+}dk_{z}\over 4\pi^{2}}
E_{l}(\eta ,k_{+},k_{z})j_{l'}(kr)P_{l}^{2}(\hat{k}\cdot\hat{z})
P_{l'}^{\pm 2}(\hat{k}\cdot\hat{z})=
$$
$$
={\bf M}_{\pm}e^{\mp 2i\phi_{\hat{r}}}\cdot
\sum_{l\ge 2}
(-i)^{l}\sqrt{(l-2)!\over (l+2)!}\cdot
\sum_{l'\ge 2}
i^{l'}(2l'+1)P_{l'}^{\mp 2}(0)P_{l'}^{\pm 2}(0)\cdot
$$
\begin{equation}
\label{cylqu11}
\cdot
\int{k_{+}dk_{+}dk_{z}\over 4\pi^{2}}
E_{l}(\eta ,k_{+},k_{z})e^{ik_{z}z}j_{l'}(k_{+}r_{+})
P_{l}^{2}(\hat{k}\cdot\hat{z})
\ \ ,\ \ ({\rm valid\ for}\ \hat{n}=\hat{z})\ ,
\end{equation}
$$
(Q\pm iU){\bf M}_{\pm}=
{\bf M}_{\pm}e^{\mp 2i\phi_{\hat{r}}}\cdot
$$
\begin{equation}
\label{cylqu12}
\cdot
\sum_{l\ge 2}
(-i)^{l}\sqrt{(l-2)!\over (l+2)!}
\cdot\int{k_{+}dk_{+}dk_{z}\over (2\pi )^{3}}e^{ik_{z}z}
E_{l}(\eta ,k_{+},k_{z})P_{l}^{2}(\hat{k}\cdot\hat{z})
{\bf JE}_{+}(\pm 2,k_{+}r_{+})\ \ , \ \ 
({\rm valid\ for}\ \hat{n}=\hat{z})\ .
\end{equation}
This corresponds to the case sketched in the upper panel of 
figure 2, where the propagation direction is parallel to the 
symmetry axis. 
As in the spherical case, orienting the $lab$-axis as in the figure 
(so that $\phi_{\hat{r}}=0$) yields an equal contribution from the 
$\pm$ terms; the polarization is given by a pure $Q$ 
term, giving the difference between the temperature fluctuations 
of the light polarized in the directions shown in the upper panel 
of figure 2; also, $\alpha =0$ in (\ref{pd}), meaning that the 
polarization direction lies on the plane formed by $\hat{n}$ and 
$\hat{z}$ (and it is orthogonal to $\hat{n}$ of course). 
The same quantities for an infinite cylindrical seed are 
easily gained using the Dirac delta 
(the sum is restricted to even $l$ and $l'$ from 
(\ref{legendre0m})):
$$
(Q\pm iU){\bf M}_{\pm}=
{\bf M}_{\pm}e^{\mp 2i\phi_{\hat{r}}}\cdot
\sum_{l\ge 2}
\sqrt{(l-2)!\over (l+2)!}P_{l}^{2}(0)\cdot
\sum_{l'\ge 2}
(-1)^{(l+l')/2}(2l'+1)P_{l'}^{\mp 2}(0)P_{l'}^{\pm 2}(0)\cdot
$$
\begin{equation}
\label{cyliqu11}
\cdot
\int{k_{+}dk_{+}\over 2\pi}
E_{l}(\eta ,k_{+})j_{l'}(k_{+}r_{+})
\ \ ,\ \ ({\rm valid\ for}\ \hat{n}=\hat{z}\ {\rm even}\ l,l')\ ,
\end{equation}
$$
(Q\pm iU){\bf M}_{\pm}=
{\bf M}_{\pm}e^{\mp 2i\phi_{\hat{r}}}\cdot
$$
\begin{equation}
\label{cyliqu12}
\cdot
\sum_{l\ge 2}
(-1)^{l/2}\sqrt{(l-2)!\over (l+2)!}P_{l}^{2}(0)
\cdot\int{k_{+}dk_{+}\over 4\pi^{2}}
E_{l}(\eta ,k_{+}){\bf JE}_{+}(\pm 2,k_{+}r_{+})\ \ , \ \ 
({\rm valid\ for}\ \hat{n}=\hat{z},\ {\rm even}\ l)\ .
\end{equation}
Just like the spherical case, photons propagating exactly 
{\it on} the symmetry axis have 
to be not polarized, since no preferred axis 
exists on the polarization plane. 
Let's check that the above results are 
consistent with this geometric expectation. 
In equations (\ref{cylqu11}) and (\ref{cyliqu11})
this is manifest because the only Bessel 
function that would survive on the axis 
($r_{+}=0$) would be $j_{0}$,  
but it's not present, since $l'\ge 2$. 
For what concerns equations (\ref{cylqu12}) 
and (\ref{cyliqu12}), the ${\bf JE}_{+}$ function 
in $r_{+}=0$ is trivially $0$ as is 
evident from (\ref{useful}).

There is another case of interest for an infinite cylindrical 
structure: precisely when the $\hat{n}$ direction is orthogonal 
to the axis. In this case the polarization plane and the equatorial 
plane are orthogonal; than it's easy to see that 
$\alpha_{\hat{k}}$ is $\tilde{\alpha}$ 
or $\tilde{\alpha}+\pi$ where the constant $\tilde{\alpha}$ 
is simply the angular 
coordinate of the projection of the equatorial 
plane into the polarization one: 
this is simply because, for the effect of 
the Dirac delta, the integration is confined 
into the equatorial plane $k_{z}=0$. 
A necessary step here is to use a note expansion 
of the second order Legendre polynomial 
in term of the elementary ones
\begin{equation}
\label{note}
P_{l}^{2}(\hat{k}\cdot\hat{n})=
\sum_{j\le l}
a_{jl}P_{j}(\hat{k}\cdot\hat{n})\ ,
\end{equation}
where the coefficients $a_{jl}$ are defined in appendix B, equation 
(\ref{numerical}). The arguments widely applied in this section 
lead to the following expressions of this interesting case: 
$$
(Q\pm iU){\bf M}_{\pm}=
{\bf M}_{\pm}e^{\mp 2i\tilde{\alpha}}\cdot
\sum_{l\ge 2}
\sqrt{(l-2)!\over (l+2)!}
\sum_{j,m=-j}^{j}
a_{jl}P_{j}^{m}(0)P_{j}^{-m}(0)
e^{im(\phi_{\hat{n}}-\phi_{\hat{r}})}\cdot
\sum_{l'\ge |m|}
(-1)^{l+(l'+l)/2}(2l'+1)\cdot
$$
\begin{equation}
\label{cyliquo}
\cdot P_{l'}^{m}(0)P_{l'}^{-m}(0)
\cdot\int{k_{+}dk_{+}\over 2\pi}
E_{l}(\eta ,k_{+})j_{l'}(k_{+}r_{+})\ ,\ ({\rm valid\ for}
\ \hat{n}\cdot\hat{z}=0,{\rm even}\ j+l,l+m,l'+m,l+l')\ ,
\end{equation}
$$
(Q\pm iU){\bf M}_{\pm}=
{\bf M}_{\pm}e^{\mp 2i\tilde{\alpha}}\cdot
\sum_{l\ge 2}
(-i)^{l}\sqrt{(l-2)!\over (l+2)!}
\sum_{j,m=-j}^{j}
a_{jl}P_{j}^{m}(0)P_{j}^{-m}(0)
e^{im(\phi_{\hat{n}}-\phi_{\hat{r}})}\cdot
$$
\begin{equation}
\label{cyliquawo}
\cdot\int{k_{+}dk_{+}\over 4\pi^{2}}
E_{l}(\eta ,k_{+}){\bf JE}_{+}(m,k_{+}r_{+})\ \ ,\ \ 
({\rm valid\ for}\ \hat{n}\cdot\hat{z}=0);
\end{equation} 
the restriction to the sum in (\ref{cyliquo}) 
comes from the properties of the expansion 
coefficients $a_{jl}$ in (\ref{numerical}) and 
again from (\ref{legendre0m}). Despite of the large number 
of sums, equation (\ref{cyliquo}) is 
workable because all the Legendre 
polynomials are calculated in the equatorial plane,  
$(\hat{n}\cdot\hat{z})=
(\hat{r}\cdot\hat{z})=(\hat{k}\cdot\hat{z})=0$; it will 
be used for the numerical integrations in section V. 
Both the expressions explicitly show the symmetry of the
seed; choosing the axes on the 
polarization plane parallel and orthogonal 
to the symmetry axis 
(so that $\tilde{\alpha}=0$) implies that 
(\ref{cyliquo}) and (\ref{cyliquawo}) 
give no distinction between the $\pm$ 
modes, giving again a pure 
$Q$ term; thus the polarization direction lies 
in the equatorial plane, 
as displayed in the upper panel of figure 2. 
As the very final observation, 
note that, in contrast to the case 
$\hat{n}=\hat{z}$, now photons 
propagating away from the symmetry axis 
at $r_{+}=0$ {\it can} be polarized; a numerical 
demonstration of this occurrence will be given 
in section V. Physically this is because 
there is a preferred axis on the polarization plane, 
the symmetry axis itself; formally, 
now the $m=j=0$ term is 
admitted, so that $j_{0}$ at $r_{+}=0$ in (\ref{cyliquo}) 
and ${\bf JE}_{+}=2\pi\delta_{m0}$ in (\ref{cyliquawo}) 
survive; it is straightforward to write down the polarization 
tensor in this particular case:
\begin{equation}
\label{figo}
(Q\pm iU){\bf M}_{\pm}=
{\bf M}_{\pm}e^{\mp 2i\tilde{\alpha}}\cdot
\sum_{l\ge 2}
(-1)^{l/2}\sqrt{(l-2)!\over (l+2)!}
\sum_{j\ge l}
a_{jl}[P_{j}(0)]^{2}\cdot
\int{k_{+}dk_{+}\over 2\pi}E_{l}(\eta ,k_{+})
\ \ ({\rm even}\ l,j)\ ;
\end{equation} 
it depends on nothing more, except for the nature of the 
infinite seed, encoded in the $E_{l}$ coefficients. 

The equations I have developed here and in the previous 
section describe CMB 
perturbations, both for temperature and polarization, 
around symmetric seeds at a given time specified by $\eta$. 
In the next section I show how to get their appearance 
on the CMB sky. 

\section{Polarization and temperature anisotropies}
\label{polarizationand}

The expressions in the previous sections 
describe the CMB polarization and temperature {\it perturbations}
around a symmetric structure, as a function of the conformal
time $\eta$ and the geometry of the seed itself. At any time, 
if a perfect CMB detector is placed around one of the seed analyzed, 
the measure of the CMB perturbation carried by a photon 
propagating on a direction $\hat{n}$ would give the 
appropriate result from the above formulas. 

Now let's face the computation of the CMB {\it anisotropy} from a 
symmetric seed. This requires the convolution of the CMB perturbation 
with the decoupling history of the universe. 
According to the current scenario, 
CMB photons were last scattered far from us in spacetime, 
when the scale factor was approximatively one thousandth than now. 
Such process is described by the last scattering probability 
between $\eta$ and $\eta +d\eta$, 
function of several cosmological parameters and of the time
of course; its expression in terms of the differential optical
depth $\tau (\eta)$ (see section V) is very simple:
\begin{equation}
\label{lssprob}
P(\eta )=\dot{\tau}e^{-\tau}\ .
\end{equation}
With the appropriate numbers, the last scattering probability
peaks on a spherical corona around us moving away with the light speed 
of course; it has present radius and thickness of about 
$6000h^{-1}$ and $10h^{-1}$ comoving Mpc respectively: for its 
thinness this zone is called last scattering surface (LSS). 
Since it is useful here, I recall that using the conformal time as 
temporal coordinate is also convenient since a photon last scattered at 
$\eta$ has to travel a comoving distance $\eta_{0}-\eta$ to reach our 
spacetime position, indicated in the following 
with the subscript $_{0}$. 

As mentioned in the introduction, the most 
known class of primordial perturbations is Gaussian 
and (nearly) scale-invariant; a simplification 
allowed by this statistics is that the CMB anisotropies have the 
same spectrum regardless of the position 
of the observer (Cosmic Variance 
subtracted of course \cite{CV}). 
The seeds analyzed here represent a radically 
different CMB anisotropy source; 
technically speaking they are 
non-Gaussian and non-scale-invariant. 
As a consequence of this, the {\it position} of the source 
along the photons path becomes here a physical degree of freedom, 
and the classification of the various possibilities is essential to 
predict how the CMB signal from a symmetric seed could appear. 

Let us start from the spherical symmetry. 
As sketched in the lower panel of figure 1, the CMB 
anisotropies are completely specified by the comoving 
distance $d$ between the seed center and the LSS peak: 
the latter is defined as the point from which we 
receive CMB photons with highest probability (peak of $P(\eta )$) 
on the direction $\hat{n}_{c}$ corresponding to the center of the
spherical seed; the observer is far on the right 
and receives on a direction $\hat{n}$ the CMB photons 
last scattered inside the spherical 
perturbation, with probability sketched as a Gaussian in the 
figure. The whole signal is symmetric with respect to 
rotations around $\hat{n}_{c}$. Also it is convenient to define 
the useful angle $\theta$ by
\begin{equation}
\hat{n}\cdot\hat{n}_{c}=\cos{\theta}\ ;
\end{equation}
it is simply the angle between the photon propagation direction
$\hat{n}$ and the direction corresponding to photons coming from
the center of the spherical seed in the sky. 
A photon last scattered at $\eta$ with direction 
$\hat{n}$ carries a CMB perturbation computable 
with the formulas developed in the previous section, that 
require its radial coordinate $r$; 
the latter is completely fixed by $\eta$, $d$ and $\theta$: 
\begin{equation}
\label{los}
r=[(d+\eta_{0}-\eta_{LSS})^{2}
+(\eta_{0}-\eta )^{2}-2(d+\eta_{0}-\eta_{LSS})
(\eta_{0}-\eta)\cos\theta ]^{1/2}\ ,
\end{equation}
where $\eta_{LSS}$ and $\eta_{0}$ mean LSS peak and 
present conformal times respectively; in fact, since 
$\eta_{0}-\eta$ is just the comoving causal 
distance covered by a photon
last scattered at $\eta$ and reaching us today, 
gaining equation (\ref{los})
is matter of simple trigonometry, see figure 1. 
This completes the spherically symmetric case. 
Once we have specified $d$, gaining the CMB polarization and 
temperature anisotropies from a spherical seed means performing  
line of sight integrations for each direction specified 
by $\theta$, as it is exposed below. 
Of course, the appearance on the 
sky of the CMB temperature and polarization anisotropies 
from one spherical seed is circular; more interesting, 
while nothing forbids photons coming 
on the $\hat{n}_{c}$ direction to carry a temperature perturbation, 
the geometric constraint treated in section II forces them to be not 
polarized. A nice example of this occurrence can be found in \cite{B}.

Let's face now the case of CMB anisotropies coming from 
cylindrically symmetric seeds. First of all, 
let's define the plane $\Pi$
containing the seed symmetry axis and our observation point: 
the signal is of course symmetric 
with respect to reflections on this plane. 
Also let's define a $\Pi$ orthogonal versor, $\hat{n}_{\Pi}$, and 
one along the symmetry axis, $\hat{z}$, regardless of their direction. 
Take now a representative point $\vec{C}$ on the symmetry axis; 
in the spherical case it was the sphere's center, but here, 
in principle, it could be any point 
along the axis: inside the seeds itself, 
or the axis intersection with the LSS peak, 
or ultimately the point of minimal 
distance from the observation point. 
Let's define $\hat{n}_{C}$ as the direction 
of photons coming from $\vec{C}$ and $D$ its 
comoving distance from the LSS peak 
(of course $\hat{n}_{\Pi}\cdot\hat{n}_{C}=0$); 
these simple geometric quantities are 
displayed in figure 2, bottom panel. 
Now take a photon last scattered at $\eta$ on a direction 
$\hat{n}$, described with the usual 
angles $\theta$ and $\phi$ in the frame 
defined by $\hat{e}_{3}=\hat{z}$, $\hat{e}_{1}=\hat{n}_{\Pi}$ and 
$\hat{e}_{2}=\hat{z}\times\hat{n}_{\Pi}$ (only $|\pi /2-\phi |$ 
would be necessary, since 
the signal does not change for reflections on $\Pi$, look at figure 2). 
Let's define for a moment $\vec{P}$ and $\vec{O}$ to be the photon 
scattering point and the observation point 
as seen by the frame centered in $\vec{C}$: 
\begin{equation}
\label{p}
\vec{P}=r_{+}\cos\phi \,
\hat{n}_{\Pi}+r_{+}\sin\phi 
\,\hat{z}\times\hat{n}_{\Pi}+z\,\hat{z}\ ,
\end{equation}
\begin{equation}
\label{o}
\vec{O}=(D+\eta_{LSS})\cdot\hat{n}_{C}\ ;
\end{equation}
in order to employ the equations 
developed in the previous section we need to know 
$r_{+}$ and $z$. This is easily 
done by expressing $\vec{P}$ as seen in a system 
with the same axes orientation but centered in $\vec{O}$:
\begin{equation}
\label{pp}
\vec{P}'=-(\eta_{0}-\eta )\cdot\hat{n}=\vec{P}-\vec{O}\ .
\end{equation}
This fixes the quantities needed: 
$$
r_{+}=[(\eta_{0}-\eta )^{2}\sin^{2}\theta +
(D+\eta_{LSS})^{2}[(\hat{n}_{C}\cdot\hat{z}\times\hat{n}_{\Pi})^{2}
+(\hat{n}_{C}\cdot\hat{n}_{\Pi})^{2}]-
$$
\begin{equation}
\label{cylrlos}
-2\sin\theta (\hat{n}_{C}\cdot\hat{n}_{\Pi}\cos\phi +
\hat{n}_{C}\cdot\hat{z}\times\hat{n}_{\Pi}\sin\phi )
(\eta_{0}-\eta )(D+\eta_{LSS})]^{1/2}\ ,
\end{equation}
\begin{equation}
\label{cylzlos}
z=-(\eta_{0}-\eta)\cos\theta+(\eta_{LSS}+D)
\hat{n}_{C}\cdot\hat{z}\ .
\end{equation}
As expected, the cylindrical symmetry 
has introduced an angular variable 
more than the spherical case. 
The quantities $r_{+}$ and $z$ defined above 
allow to employ the formulas developed in the 
previous section to 
compute the CMB anisotropy carried by 
the photon last scattered at $\eta$ 
on the direction $\hat{n}$; 
of course, for an infinite cylindrical 
seed only the $r_{+}$ coordinate is necessary. 
While anisotropies in the spherical 
case are characterized by a circular 
imprint, here their shape may vary with  
the orientation of the symmetry axis. 
If it coincides with $\hat{n}_{C}$, 
thus including the observation point, 
the imprint is circular around it, 
and again polarization anisotropies 
are absent on the direction corresponding 
to the symmetry axis itself. 
In any other case, both polarization and temperature 
anisotropies would appear symmetric around a 
{\it line} in the sky, 
projection of the seed symmetry axis on the celestial sphere. 

Finally, the CMB anisotropies both for 
the spherical and cylindrical cases 
are obtained through a line of sight integration along the 
photon's path, convolved with the last scattering probability 
(see \cite{HW}): 
\begin{equation}
\label{dttani}
\Theta (\eta_{0},here,\hat{n})=\int_{0}^{\eta_{0}}
\left[(\Theta +\Psi )(\eta ,arg,\hat{n})P(\eta) +
(\dot{\Psi}-\dot{\Phi})(\eta ,arg,\hat{n})e^{-\tau}\right]d\eta\ ,
\end{equation}
\begin{equation}
\label{quani}
(Q\pm iU)(\eta_{0},here,\hat{n}){\bf M}_{\pm}=\int_{0}^{\eta_{0}}
(Q\pm iU)(\eta ,arg,\hat{n}){\bf M}_{\pm}P(\eta)d\eta\ ;
\end{equation}
at each $\eta$, (\ref{los}) and 
(\ref{cylrlos},\ref{cylzlos}) give the 
necessary arguments ($arg$) to compute 
the CMB perturbations. $\Psi$ accounts for the Sachs-Wolfe effect, 
due to the work spent by the photon climbing out of the 
potential well (or hill) in which 
it was last scattered; the time derivatives 
account for the integrated 
Sachs-Wolfe effect, due to 
the work spent by the same photon 
crossing the density perturbations 
on the way toward us. 

The following consideration introduces to the next section. 
As I have already mentioned, a symmetric seed 
could be a spatially limited 
structure, say a monopole or a 
bubble for the spherical case, or a 
string for the cylindrical case. 
Thus also the CMB anisotropy is spatially 
limited, since the evolution equations may 
transport the CMB perturbation at most 
at a sound horizon distance from the source. 
Therefore, if the perturbed zone does not 
intersects the LSS, 
meaning that it occupies 
a spacetime region where $P(\eta )$ is 
negligibly small, the terms $\Theta$, 
$Q$ and $U$ above do not 
give contributions; 
in this situation, the seed can't signal its presence, 
except for the integrated Sachs-Wolfe effect if it lies 
within our Hubble sphere (a distinctive and fascinating 
signal in this case arise from cosmic strings \cite{BBS}). 
Thus, in order to detect the genuine CMB signal from a 
symmetric spatially limited seed, we should be lucky with its 
spacetime location: it should intersect the LSS. 

\section{The pebbles in a pond}

Let us apply the formulas developed in the previous sections.
I plunge a toy symmetric source in the cosmic fluid
at the initial time $\eta =0$, computing its evolution by using
the linear theory of the cosmological perturbations. 
At different times during the evolution, some pictures of the
corresponding CMB polarization and temperature perturbations
are taken. Finally, the computation of the line of sight integrals
(\ref{dttani}) and (\ref{quani}) simulates the CMB signal as
it would appear in an high resolution observation.

First, let us define the initial density perturbations.
For the spherical case, I take a potential energy 
condensation with a Gaussian shape extending on a 
comoving radial distance $R$: 
\begin{equation}
\label{toysphere}
\Psi (r,\eta =0)={\cal N}
\exp\left[-\left({r\over R}\right)^{2}\right]\ .
\end{equation} 
For the cylindrical case, I take an infinitely long seed, 
with a potential energy condensation 
on the equatorial plane characterized again by a Gaussian 
shape and extending on a scale $R_{+}$: 
\begin{equation}
\label{toycylinder}
\Psi (r_{+},\eta=0)={\cal M}
\exp\left[-\left({r_{+}\over R_{+}}\right)^{2}\right]\ .
\end{equation}
The normalization constants will be fixed below. 
The Fourier transforms are easily performed
in the frame with origin in the center of the sources: 
\begin{equation}
\label{toyspherek}
\Psi (k,\eta =0)={\cal N}\pi^{3/2}R^{3}
\exp\left[-\left({kR\over 2}\right)^{2}\right]\ ,
\end{equation}
\begin{equation}
\label{toycylinderk}
\Psi (k_{+},\eta =0)={\cal M}\pi R_{+}^{2}
\exp\left[-\left({k_{+}R_{+}\over 2}\right)^{2}\right]\ .
\end{equation}
The evolution equations for fluid and CMB quantities may
be obtained by the Boltzmann and linearized Einstein equations
\cite{MB,BKS,HW}. A standard CDM scenario is assumed, including
cold dark matter ($_{c}$), baryons ($_{b}$), photons ($_{\gamma}$) 
and three families of massless neutrinos ($_{\nu}$). 
All the equations in the following are written in Fourier 
space. The equations for the matter species are: 
\begin{equation}
\label{c}
\dot{\delta}_{c}=-kv_{c}-3k^{2}\dot{\Phi}
\ ,\ \dot{v}_{c}=-{\dot{a}\over a}v_{c}+k\Psi\ ,
\end{equation}
\begin{equation}
\label{b}
\dot{\delta}_{b}=-kv_{b}-3k^{2}\dot{\Phi}
\ ,\ \dot{v}_{b}=-{\dot{a}\over a}v_{b}+k\Psi +
{4\rho_{\gamma}\over 3\rho_{b}}an_{e}\sigma_{T}
(v_{\gamma}-v_{b})\ ,
\end{equation}
where $\delta\equiv (\delta\rho /\rho)$ and
$\dot{\tau}=ax_{e}n_{e}\sigma_{T}$ is the differential
optical depth; $n_{e}$ is the electron 
number density and $x_{e}$ is the ionization fraction 
(see the last work in reference \cite{HW} for useful fitting 
formulas). Photon equations involve each multipole
in the expansions (\ref{dttdeveloped}) and 
(\ref{qudeveloped}):
\begin{equation}
\label{theta01}
\dot{\Theta}_{0}=-{k\over 3}\Theta_{1}-\dot{\Phi}
\ ,\ 
\dot{\Theta}_{1}=k\Theta_{0}-{2\over 5}k\Theta_{2}+
\dot{\tau}(v_{b}-\Theta_{1})+k\Psi\ ,
\end{equation}
\begin{equation}
\label{theta2e2}
\dot{\Theta}_{2} = {2\over 3}k\Theta_{1}-{3\over 7}k\Theta_{3}-
\dot{\tau}\left({9\over 10}\Theta_{2}-{\sqrt{6}\over 10}E_{2}\right)
\ ,\ 
\dot{E}_{2}=-{\sqrt{5}\over 7}kE_{3}-
\dot{\tau}\left({1\over 10}\Theta_{2}+{2\over 5}E_{2}\right)\ ,
\end{equation}
and for $l\ge 3$:
\begin{equation}
\label{dttqulboltzmann}
\dot{\Theta}_{l}=k\left[{l\over 2l-1}\Theta_{l-1}-
{l+1\over 2l+3}\Theta_{l+1}\right]-
\dot{\tau}\Theta_{l}
\ ,\ \dot{E}_{l} = k
\left[{\sqrt{l^{2}-4}\over 2l-1}E_{l-1}-
{\sqrt{(l+1)^{2}-4}\over 2l+3}E_{l+1}\right]-
\dot{\tau}E_{l}\ .
\end{equation}
In Newtonian gauge the lowest multipoles
are linked to the photon fluid quantities by
$\delta_{\gamma}=4\Theta_{0}$, $v_{\gamma}=\Theta_{1}$ and
$\pi_{\gamma}=12\Theta_{2}/5$. Massless neutrinos
can be treated as photons without the polarization and
Thomson scattering terms. Finally, the equations 
for the gravitational potentials are:
\begin{equation}
\label{phi}
k^{2}\Phi =4\pi Ga^{2}\left[\rho_{c}\delta_{c}+
\rho_{b}\delta_{b}+\rho_{\gamma}\delta_{\gamma}+
\rho_{\nu}\delta_{\nu}+{3\over k}{\dot{a}\over a}\left(
\rho_{c}v_{c}+\rho_{b}v_{b}+{4\over 3}\rho_{\gamma}v_{\gamma}
{4\over 3}\rho_{\nu}v_{\nu}\right)\right]\ ,
\end{equation}
\begin{equation}
\label{psi}
-k^{2}(\Psi +\Phi)={8\pi G\over 3}
\left(\rho_{\gamma}\pi_{\gamma}+\rho_{\nu}\pi_{\nu}\right)\ .
\end{equation}
As it is known \cite{MB,BKS}, at early times the above system
can be solved by using the tight coupling approximation
between photons and baryons. The multipole equations
are expanded in powers of $k/\dot{\tau}\ll 1$. The only
zero order terms are $\Theta_{0}$ and $\Theta_{1}$ from
(\ref{b},\ref{theta01}), and obey the following equations:
\begin{equation}
\label{thetazerothetaone}
\dot{\Theta}_{0}=-{k\over 3}\Theta_{1}-\dot{\Phi}
\ \ ,\ \ {d\over d\eta}\left[\left(1+
{3\rho_{b}\over 4\rho_{\gamma}}\right)\Theta_{1}\right]=
k\Theta_{0}+k\left(1+{3\rho_{b}\over 4\rho_{\gamma}}\right)\Psi\ \ ,
\end{equation}
where $\Theta_{1}$ is assumed to concide with $v_{b}$ to the 
lowest order. Increasing the order in $k/\dot{\tau}$
the higher multipoles are given by
\begin{eqnarray}
\label{tc2}
\Theta_{2}={k\over\dot{\tau}}{8\over 9}\Theta_{1}\ &,&\ 
E_{2}=-{\sqrt{6}\over 4}\Theta_{2}\ ,\\
\label{tcl}
\Theta_{l}={k\over\dot{\tau}}{l\over 2l-1}\Theta_{l-1}\ &,&\ 
E_{l}={k\over\dot{\tau}}{\sqrt{l^{2}-4}\over 2l-1}E_{l-1}\ .
\end{eqnarray}
I integrate in time the system (\ref{thetazerothetaone},
\ref{tc2},\ref{tcl}) until $k/\dot{\tau}=.1$ 
occurs, thereafter integrating the complete equations; 
of course, care is taken that the results do not depend at all 
on this choice. 

I take adiabatic initial conditions: at early times 
$\delta_{c}=\delta_{b}=3\delta_{\gamma}/4=3\delta_{\nu}/4$ 
(all the velocity are initially zero) and 
the second member in equation (\ref{phi}) at $\eta =0$ 
is proportional for each Fourier mode to the initial 
perturbation spectrum (\ref{toyspherek}) or 
(\ref{toycylinderk}); in order 
to make the following results more clear, the latter 
is normalized with the density contrast $\delta$ taken in 
the center of the seed at decoupling. 
This choice is not dependent on the 
particular gauge chosen here, since equation (\ref{phi}) 
is gauge invariant \cite{BKS}. In the CMB equations, 
everything is initially zero except for the lowest 
multipole of the temperature perturbation \cite{MB}: 
\begin{equation}
\label{adi}
\Theta_{0}(0)=-2\Psi(0)\ .
\end{equation}
The background evolution is driven by the Einstein equation
\begin{equation}
{\dot{a}^{2}\over a^{2}}={8\pi G\over 3}a^{2}\sum_{a}\rho_{a}\ ,
\end{equation}
where the index $a$ runs over all the fluid species. 
Now the computation system is ready. The background parameters
describe a standard CDM model ($\Omega_{0}=1,h=.5,\Omega_{b}=0.05,
\Omega_{CDM}=1-\Omega_{b}$).

Figure 3 shows the time evolution of the
CMB temperature perturbation from the spherical seed.
The chosen comoving size is $R= 10h^{-1}$ Mpc, that is
well below the effective horizon at decoupling, 
(approximatively $100h^{-1}$ Mpc).
The radial profile is shown, and the temperature
perturbation has been computed from 
equation (\ref{thetasphere}) for photons propagating
perpendicularly to the radial direction, as indicated. 
The sum converges very rapidly: the heavy line shows the 
result from the first ten multipoles, while the light one, 
almost indistinguishable, indicates the result from 
the only $l=0$ multipole in (\ref{thetasphere}). 
The figure points out the wave-like behavior of the CMB
perturbations. The initial condition (panel $a$) remains
unchanged until the horizon crossing, that occurs nearly
at equivalence for the chosen size. At this time, in 
panel $b$, baryons tend to fall into the potential
well, and the perturbation amplitude grows.
After that, in panel $c$, an opposite oscillation due to
the pressure reaction takes place, pushing the perturbation
away from the center. Finally, in panel $d$ the perturbation
is shown just before decoupling: the oscillatory
behavior caused a temperature
perturbation wave that is propagating outward. The
wave crest is just at the position of the sound horizon
at the time displayed. This phenomenology is analogous 
in figure 4, where the polarization amplitude, computed 
using equation (\ref{intpol}), is shown. 
At the initial time no perturbation is 
visible, since all the Fourier modes are outside the 
horizon. At the horizon crossing the oscillations begin, 
producing a well visible polarization wave that travels
outward with the CMB sound velocity. Note that, as an
important distinction with respect to the temperature case,
for the polarization there is no perturbation
near the center, at small $r$. This is a practical
realization of the geometric constraint exposed in
section II: photons propagating radially in a spherical
density field must be not polarized, since no preferred
axis exists on the polarization plane. 

Figures 5 and 6 show the same analysis on the cylindrical 
seed and remarkably the same undulatory 
phenomenology of the spherical case occurs. 
The size is $R_{+}=10h^{-1}$ Mpc 
and photons propagating perpendicularly to the 
symmetry and radial directions are considered, from 
equations (\ref{cylit}) and (\ref{cyliquo}). 
Again the sum converges very rapidly: the light 
line in figure 5 corresponds to the $l=0$ terms in 
(\ref{cylit}). At the horizon crossing, the competition 
between pressure and gravity generate CMB 
temperature and polarization waves propagating 
away from the symmetry axis. Just before decoupling, 
panels $d$, temperature and polarization waves 
are well visible a CMB sound horizon away from the 
axis of the cylindrical seed.  
As an interesting feature, note how in this case 
the polarization for photons scattered on the symmetry axis 
is non-vanishing: this is evident particularly in panel $c$. 
The central polarization amplitude is in any case smaller than 
the mean signal size, since for $r_{+}\rightarrow 0$ in 
(\ref{cyliquo}) only the $l=0$ term survive. 

Figure 7 shows the results of the line of sight integrals
(\ref{dttani}) and (\ref{quani}), where for simplicity 
only the spherical case is shown; I recall that $\theta$ 
is the angle between the line of sight and the one corresponding 
to the center of the seed. The importance of 
the different positions of the seed with respect to 
the LSS is evident: the solid line shows the signal if 
the spherical perturbation lies exactly on the last scattering 
surface, $d=0$, while the dashed and dotted dashed 
lines corresponds to the cases $d=30h^{-1}$ Mpc and $d=-30h^{-1}$ Mpc 
respectively. The general features pointed of 
the time evolution have been preserved. 
Simply, the CMB temperature and polarization waves 
propagating outward from the spherical seed have been 
snapped by the decoupling photons. The anisotropy waves 
extend on the scale of a CMB sound horizon at
decoupling, that is roughly $1^{o}$ in the sky. The
temperature perturbation contains a central spot,
that is absent in the polarization case. 

It is important to point out the following considerations. 
First, note that the mean amplitude of the signal follows 
the known expectations \cite{P} 
for a linear structure with size $L\le H^{-1}$ 
and density contrast $\delta$ at decoupling: 
$\delta T/T\simeq \delta (L/H^{-1})^{2}$, 
roughly ten times stronger than the polarization signal. 
From the point of view of the dark matter distribution, 
the seed lies in the very central part of the graph, 
say $\theta\le 10'$ (corresponding to less then $10h^{-1}$ Mpc 
in figure 3 and 4). Also the amplitude 
of the waves has the same mean magnitude of the signal 
coming from the location of the seed; 
really, in the polarization from a spherical seed they are 
the very dominant component of the anisotropy. 
Thus they must be considered in any simulation aiming at the 
detection of this kind of signals. 
Also they could play some 
role in the structure formation around the seed, since 
they are physically made of photons and baryons. 
Besides, from an experimental point of view this 
undulatory occurrence could help the detection if 
structure like these ones should really exist. 
Indeed the CMB signal from a spatially limited seed is
extended on the scale of a sound horizon at decoupling 
even if the size of the seed itself is smaller; 
therefore it appears as a series of
sub-degree rings centered on the position of the seed; 
this can help to discriminate between the signals 
from point astrophysical sources from genuine 
cosmological seeds of primordial origin. 
Also, as it is evident from figure 7, a marked correlation 
exists between the temperature and polarization signals. 
Of course, this would improve the signal to noise ratio 
for high resolution instruments like Planck capable 
to detect both polarization and temperature anisotropy. 

\section{Conclusion}
\label{conclusion}

At the present time, very high energy physics is still rather 
unknown and only theoretically approached. 
The breaking of high energy symmetries in the early universe 
may have left some traces of their occurrence, like 
topological defects or true vacuum bubbles. These relics act as 
seeds for polarization and temperature 
anisotropies in the cosmic microwave 
background (CMB), and this work aims at 
providing a general framework in order 
to predict their signal. 

I have considered the cases of spherical 
and cylindrical symmetry of the 
perturbation source; no other specification 
characterizes the seed. 
I have obtained general formulas describing CMB 
polarization and temperature perturbations, 
as a function of time, generated by the most general 
structures characterized by the mentioned symmetries. 
The analysis regards both 
the pure CMB perturbation 
nearby the seeds and their CMB 
anisotropy as observed in our sky. 
Such expressions explicitly show several nice features 
to their own CMB imprint. 
 
In spherical symmetry, the polarization and temperature 
perturbations depend geometrically 
on the scalar product 
$\hat{n}\cdot\hat{r}$, where the 
first is the photon propagation 
direction and the second the radial versor in the point 
where CMB is being measured. I give explicit 
expressions in which this dependence 
is factored out of the integral over the Fourier 
perturbations modes. 
In particular the polarization direction 
(orthogonal to $\hat{n}$ of course) 
lies on the plane formed by $\hat{n}$ and $\hat{r}$. 
As an important difference between polarization and 
temperature perturbations, 
the light propagating from the center of the seed is 
not polarized, since the radial propagation in spherical 
symmetry is an axial symmetric problem, 
so that no preferred axis exists 
for the polarization; instead nothing forbids 
a temperature perturbation.

In cylindrical symmetry the polarization and 
temperature perturbations depend on the products 
$\hat{n}\cdot\hat{z}$ and 
$\hat{r}\cdot\hat{z}$, where $\hat{z}$ is 
the symmetry axis, as well as on the 
angular difference between the 
projections of $\hat{n}$ and 
$\hat{r}$ on the plane orthogonal to 
$\hat{z}$; the $\hat{r}\cdot\hat{z}$ 
dependence is lost if the seed is 
invariant for traslations along the 
symmetry axis (mentioned as 
infinite in the following). 
I give formal expressions showing these dependences, and 
extract them analytically from the Fourier integral 
in the cases of propagation parallel and 
orthogonal to the symmetry axis. 
In the first case the polarization 
direction lies on the plane formed by 
$\hat{n}$ and $\hat{z}$; as for the 
spherical case, photons traveling exactly on 
the symmetry axis are not polarized. 
In the second case, and for an infinite seed, 
the polarization direction is orthogonal to the symmetry axis.  

For what concerns the CMB anisotropies as observed in our 
sky, they are computed with an usual line of sight integration, but 
the seeds considered here introduce additional variables 
with respect to the ordinary Gaussian case, that specify their 
position and orientation along the photons path toward us, 
characterizing their appearance on our CMB sky. 

Polarization and temperature anisotropies 
from a spherical seed are circular and 
specified by the distance $d$ between the seed 
center and the LSS peak. As a consequence 
of the geometric constraints 
summarized above, CMB polarization 
anisotropy is absent for photons coming 
from the center of the seed; on the other 
hand, nothing prevents them to possess 
a temperature perturbation. 

Anisotropies from a cylindrical 
seed are specified by the distance $D$ between 
a representative point on the symmetry axis 
and the LSS peak, as well as 
on the angular orientation of 
the symmetry axis itself on the plane containing it 
and the observation point. 
Anisotropies may appear in different ways. 
If the symmetry axis includes the 
observation point, what we would see is 
a circular imprint again; as in the 
spherical case, CMB polarization anisotropy 
is absent for photons coming from the center. 
In any other case, anisotropies 
would appear symmetric around a line in the sky, 
projection of the axis on the 
celestial sphere, thus giving the 
genuine sign of a cylindrical seed. 

I have performed some numerical work on the 
formulas developed here, adopting toy symmetric sources 
in order to see the pure CMB processes at work with this 
kind of seed. The time evolution of the seed and of its 
corresponding CMB perturbation is performed from the initial 
time, and several pictures are taken before decoupling. 
The integrations highlight the undulatory behavior of 
the CMB perturbations. Just like a pebble in a pond, 
the oscillations occurring at the horizon crossing 
produce temperature and polarization perturbation 
waves that propagate outward with the CMB sound velocity. 
Consequently, the CMB anisotropies caused from structures 
like the ones analyzed here that intersect the last 
scattering surface 
extend at least on $1^{o}$ in the sky, that is the angular 
scale corresponding to the CMB sound horizon at decoupling; 
the signals contain anisotropy waves, each one characterized 
by its own value of temperature and polarization perturbation. 
This component of the signal possesses the same magnitude 
of the one coming directly from the seed interior. 
The mean amplitude roughly follow the known expectations 
for a linear structure with size $L\le H^{-1}$ 
and density contrast $\delta$ at decoupling: 
$\delta T/T\simeq \delta (L/H^{-1})^{2}$, 
roughly ten times stronger than the polarization signal, 
where $H^{-1}$ is the size of the Hubble length at decoupling. 
The anisotropy waves coming out of a symmetric 
spatially limited seed are a unique proof that the 
seed itself existed well before decoupling; thus, 
these waves could allow to distinguish relics 
from high energy processes of the early universe from point-like 
astrophysical sources, because of the angular extension and 
amplitude. Also, this phenomenology offers cross correlation 
possibilities for detectors like Planck capable to explore both 
temperature and polarization CMB sky. 

Future works will deal with models of real symmetric structures, 
relics from high energy physics. These works aim at 
predicting their appearance on the CMB map itself before 
than on the anisotropy power spectrum. 
Their detection in the high resolution CMB maps 
provided by the Microwave Anisotropy 
Probe and Planck missions in the 
next decade would be an invaluable insight into the
hidden sector of high energy physics.

\acknowledgments

The first half of this work was performed at 
the NASA/Fermilab Astrophysics center. 
It was supported by the DOE and the NASA grant NAG 5-7092. 
I warmly thank the hospitality of the Theoretical Astrophysics Group. 
Also I wish to thank Luca Amendola and Franco Occhionero for 
constant encouragement. Finally, I am grateful to Luigi Danese 
and the astrophysics sector at SISSA/ISAS, where this work 
was completed. 

\appendix

\section{Spherical harmonics and related quantities}
\label{sphericalharmonics}

The spherical harmonics are expressed as usual as 
\begin{equation}
\label{sh}
Y_{l}^{m}(\theta ,\phi)=
\sqrt{{(2l+1)\over 4\pi}{(l-m)!\over (l+m)!}}
P_{l}^{m}(\cos\theta )e^{im\phi}\ ,
\end{equation}
where the Legendre polynomials are defined by
\begin{equation}
\label{plm}
P_{l}^{m}(x)=(-1)^{m}(1-x^{2})^{m/2}
{d^{m}\over dx^{m}}P_{l}(x)\ \ ,
\ \ P_{l}^{-m}(x)=(-1)^{m}{(l-m)!\over (l+m)!}P_{l}^{m}(x)
\ \ ,\ (m\ge 0)\ , 
\end{equation}
where $x=\cos\theta$. 
Legendre polynomials and spherical harmonics obey the 
orthogonality relations
\begin{equation}
\label{orthoplm}
\int_{-1}^{1}dx\ P_{l}^{m}(x)P_{l'}^{m}(x)=
\delta_{ll'}{2\over 2l+1}{(l+m)!\over (l-m)!}\ \ ,\ \ 
\int \sin\theta d\theta d\phi Y_{l}^{m*}(\theta,\phi )
Y_{l'}^{m'}(\theta,\phi )=\delta_{ll'}\delta_{mm'}\ ,
\end{equation}
and are eigenmodes of the parity operator: 
\begin{equation}
\label{parity}
P_{l}^{m}(-x)\rightarrow (-1)^{l+m}P_{l}^{m}(x)\ .
\end{equation}
Legendre polynomials satisfy the following 
note recurrence relations:
\begin{equation}
\label{rec}
(l-m)P_{l}^{m}(x)=x(2l-1)P_{l-1}^{m}(x)-(l+m-1)P_{l-2}^{m}(x)\ ,
\end{equation}
\begin{equation}
\label{known}
P_{l}^{m+2}(x)+{2(m+1)x\over \sqrt{1-x^{2}}}P_{l}^{m+1}(x)+
(l-m)(l+m+1)P_{l}^{m}(x)=0\ ,
\end{equation}
\begin{equation}
\label{der}
(1-x^{2}){dP_{l}^{m}\over dx}=-lxP_{l}^{m}(x)+(l+m)P_{l-1}^{m}(x)\ ;
\end{equation}
they can be used to gain the value of any Legendre polynomials 
in $x=0$: 
$$
P_{0}^{0}(0)=1\ ,\ P_{1}^{0}(0)=0\ ,
\ P_{1}^{1}(0)=-1\ ,\ P_{2}^{1}(0)=0\ ,
$$
\begin{equation}
\label{legendre0}
(l-m)P_{l}^{m}(0)=-(l+m-1)P_{l-2}^{m}(0)\ \ ,\ \ P_{l}^{m+2}(0)=
-(l-m)(l+m+1)P_{l}^{m}(0)\ .
\end{equation}
Also note that 
\begin{equation}
\label{legendre0m}
P_{l}^{m}(0)=0\ \ {\rm for\ odd}\ l+m\ .
\end{equation}
In this work I have often used the addition relation for 
spherical harmonics, given by 
\begin{equation}
\label{addition}
P_{l}(\hat{n}_{1}\cdot\hat{n_{2}})={4\pi\over 2l+1}
\sum_{m=-l}^{l}
Y_{l}^{m}(\hat{n}_{1})Y_{l}^{m*}(\hat{n}_{2})\ ,
\end{equation}
and the following useful integral relation, 
that may be verified 
easily using the addition relation itself:
\begin{equation}
\label{legendre}
\int d\Omega_{\hat{n}_{1}}P_{l}(\hat{n}_{1}\cdot\hat{n}_{2})
P_{l'}(\hat{n}_{1}\cdot\hat{n}_{3})=\delta_{l l'}
{4\pi\over 2l+1}P_{l}(\hat{n}_{2}\cdot\hat{n}_{3})\ .
\end{equation}
Second order Legendre polynomials admit 
the following expansion \cite{C}:
\begin{equation}
\label{numerical}
P_{l}^{2}(x)=
\sum_{j\le l}
a_{jl}P_{j}(x)\ ,\ {\rm where}
\end{equation}
\begin{eqnarray}
a_{jl}&=& 0\ {\rm for}\ j>l\ {\rm or}\ l+j\ {\rm odd}\ ,\nonumber\\
a_{jl}&=&-2l(l-1)(2j+1)/(4l+2)\ {\rm for}\ l=j\ ,\\
a_{jl}&=& 2(2j+1)\ {\rm for}\ j<l\ {\rm and}\ l+j\ {\rm even}
\nonumber\ .
\end{eqnarray}
The tensor spherical harmonics are defined in terms 
of the ordinary ones by 
\begin{equation}
\label{tsh}
{ }_{\pm 2}Y_{l}^{m}(x)=
\sqrt{(l-2)!\over (l+2)!}\left[\partial_{\theta}^{2}-
{\rm cot}\theta\partial_{\theta}\mp {2m\over \sin\theta}
\left(\partial_{\theta}-{\rm cot}\theta\right)+
{m^{2}\over\sin^{2}\theta}\right]Y_{l}^{m}(\theta,\phi )\ ;
\end{equation}
the normalization coefficient may vary in literature. 
Equation (\ref{tensorlegendre}) is easily obtained 
in the following way. 
From (\ref{known}) one can immediately sees that 
\begin{equation}
\label{step}
{\rm cot}\theta\partial_{\theta}P_{l}(x)=
-{1\over 2}P_{l}^{2}(x)-
{l(l+1)\over 2}P_{l}(x)\ .
\end{equation}
Also the equality
\begin{equation}
\label{derirec}
\partial_{\theta}^{2}P_{l}(x)=
{x\over\sqrt{1-x^{2}}}P_{l}^{1}(x)+
P_{l}^{2}(x)
\end{equation}
holds by using elementary derivation. 
Using again (\ref{known}) for 
$m=0$ and putting (\ref{step}) and (\ref{der}) 
together, the wanted equation is obtained: 
\begin{equation}
\label{tensorlegendreapp}
_{2}Y_{l}^{0}(x)=
\sqrt{{2l+1\over 4\pi}{(l-2)!\over (l+2)!}}
P_{l}^{2}(x)\ .
\end{equation}

\section{Anger and Weber functions}
\label{angerweber}

This appendix contains some useful 
integration relations. Focus on the integral 
\begin{equation}
\label{hyptrig}
\int_{0}^{\pi}\exp[\pm i(\nu\phi-\beta\sin\phi)]d\phi=
\pi [{\bf J}_{\nu}(\beta )\pm i{\bf E}_{\nu}(\beta )]\ \ ,
\ \ ({\rm Re} \beta >0)\ , 
\end{equation}
where ${\bf J}_{\nu}$ and ${\bf E}_{\nu}$ are the Anger 
and Weber functions respectively 
(see \cite{GR} for useful recurrence relations):
\begin{equation}
\label{anger}
{\bf J}_{\nu}(z)={1\over\pi}\int_{0}^{\pi}
\cos (\nu\theta -z\sin\theta )d\theta\ ,
\end{equation}
\begin{equation}
\label{weber}
{\bf E}_{\nu}(z)={1\over\pi}\int_{0}^{\pi}\sin 
(\nu\theta -z\sin\theta )d\theta\ ,
\end{equation}
The two following equalities are easily gained using 
elementary integration algebra:
\begin{equation}
\label{eleintalg1}
\int_{0}^{2\pi}\exp [\pm i(\nu\phi-\beta\sin\phi)]d\phi=
\pi [{\bf J}_{\nu}(\beta )\pm i{\bf E}_{\nu}(\beta )]+
\pi e^{\pm i\nu\pi}
[{\bf J}_{-\nu}(\beta )\mp i{\bf E}_{-\nu}(\beta )]\ ,
\end{equation}
\begin{equation}
\label{eleintalg2}
\int_{0}^{2\pi}\exp 
[\pm i(\nu\phi-\beta\sin\phi)]d\phi=
e^{\pm i\nu\pi /2}\int_{-\pi /2}^{3\pi /2}
\exp [\mp i(-\nu\phi+\beta\cos\phi)]d\phi\ .
\end{equation}
If $\nu$ is integer, all the functions in 
the integrals are periodical on the $2\pi$ interval, so as 
the integrals above do not depend on the starting point. 
Thus the following equality holds: 
\begin{equation}
\label{useful}
\int_{0}^{2\pi}\exp [\pm i(-m\phi+\beta\cos\phi)]d\phi =
\pi e^{\pm im\pi /2}
[{\bf J}_{m}(\beta )\mp i{\bf E}_{m}(\beta )]+
\pi e^{\mp i\nu\pi/2}[{\bf J}_{-m}(\beta )\pm 
i{\bf E}_{-m}(\beta )]=
{\bf JE}_{\pm}(m,\beta )\ ;
\end{equation}
it's valid for $m=\nu$ integer and Re$\beta >0$; 
the last equality is 
a pure definition. Note that in the particular case
$\beta =0$, the above expression reduces simply to
$2\pi\delta_{m0}$.

\begin{figure}
\label{p1}
\psfig{figure=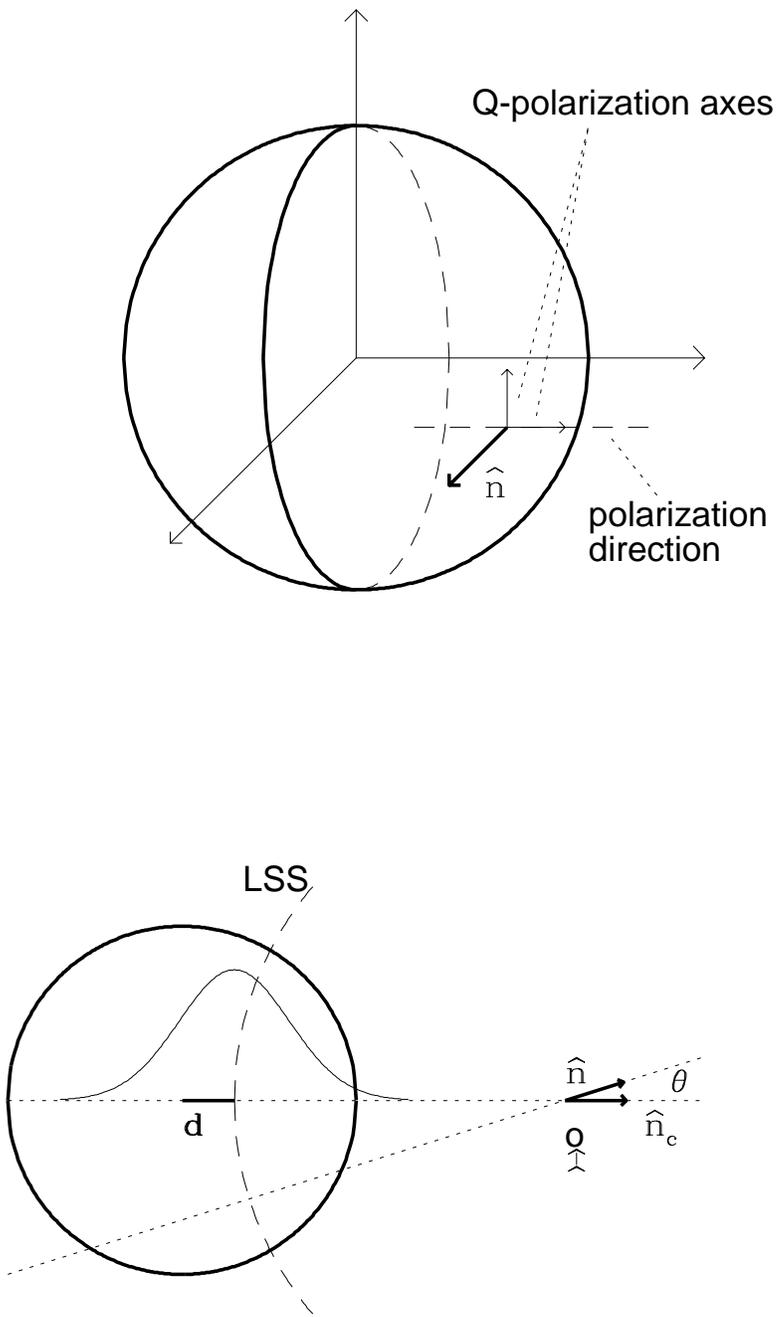}
\caption{Upper panel: 
polarization within a spherical seed. The axes displayed 
show the geometric directions for which 
the polarization is given by a $Q$ term only, 
thus fixing the polarization direction as displayed. 
Lower panel: CMB anisotropies from a spherical seed. 
Its center has a distance $d$ from 
the last scattering surface; the anisotropy 
is symmetric under rotations around 
$\hat{n}_{c}$ and depends geometrically on the angle $\theta$ only.}
\end{figure}

\begin{figure}
\label{p2}
\psfig{figure=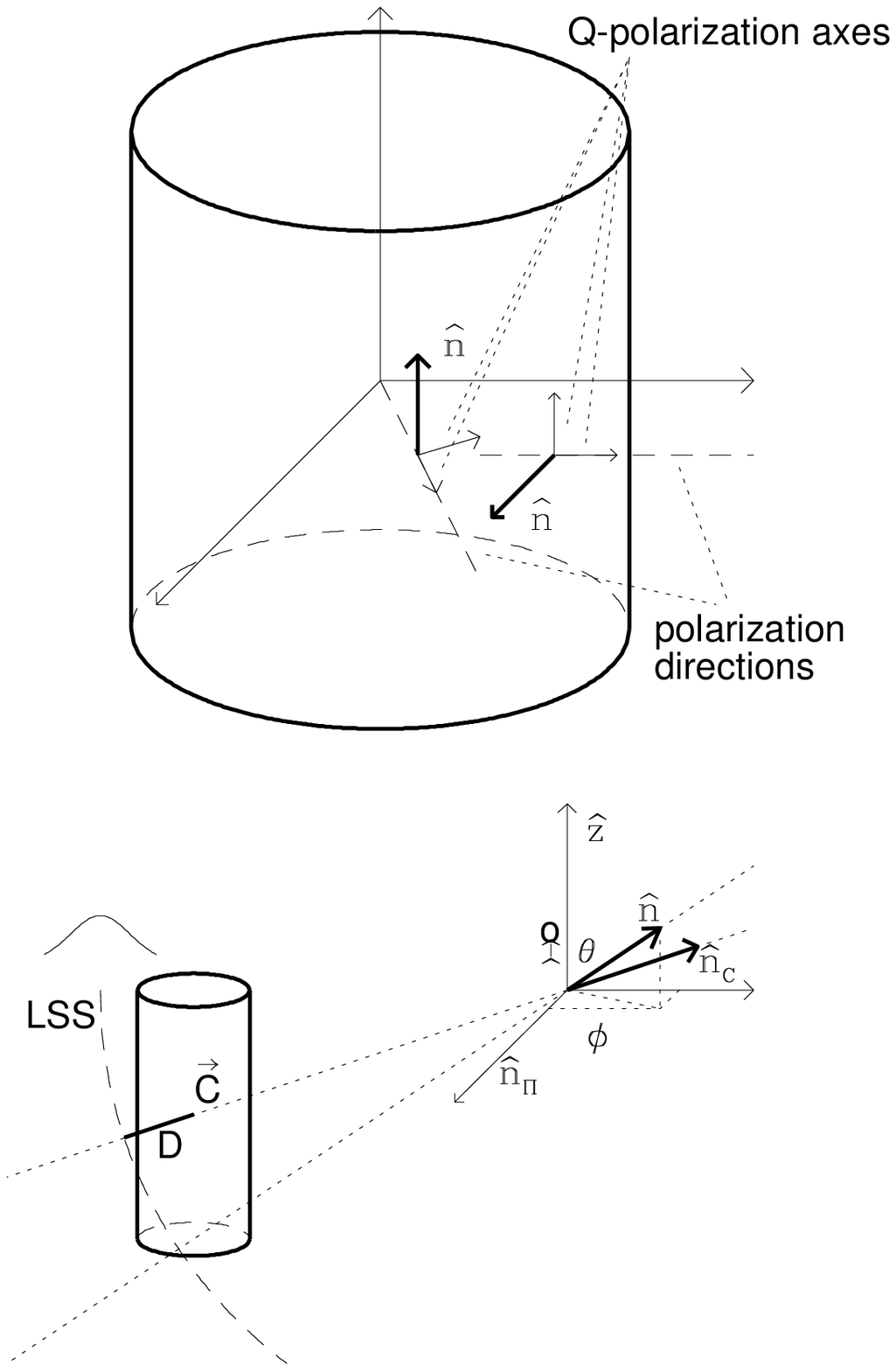}
\caption{Upper panel: polarization within a cylindrical 
seed. For the cases of propagation parallel and orthogonal 
to the symmetry axis, 
the axes displayed show the geometric 
directions for which the polarization is given 
by a $Q$ term only, thus fixing the polarization 
directions as displayed. 
Lower panel: CMB anisotropies 
from a cylindrical seed: a view of the $\Pi$ plane. 
The representative point $\vec{C}$ has a distance $D$ from 
the last scattering surface; 
the anisotropy is symmetric under reflections on $\Pi$ 
and depends geometrically 
on the angle $\theta$ and on $|\pi /2-\phi|$.}
\end{figure}

\begin{figure}
\label{p3}
\psfig{figure=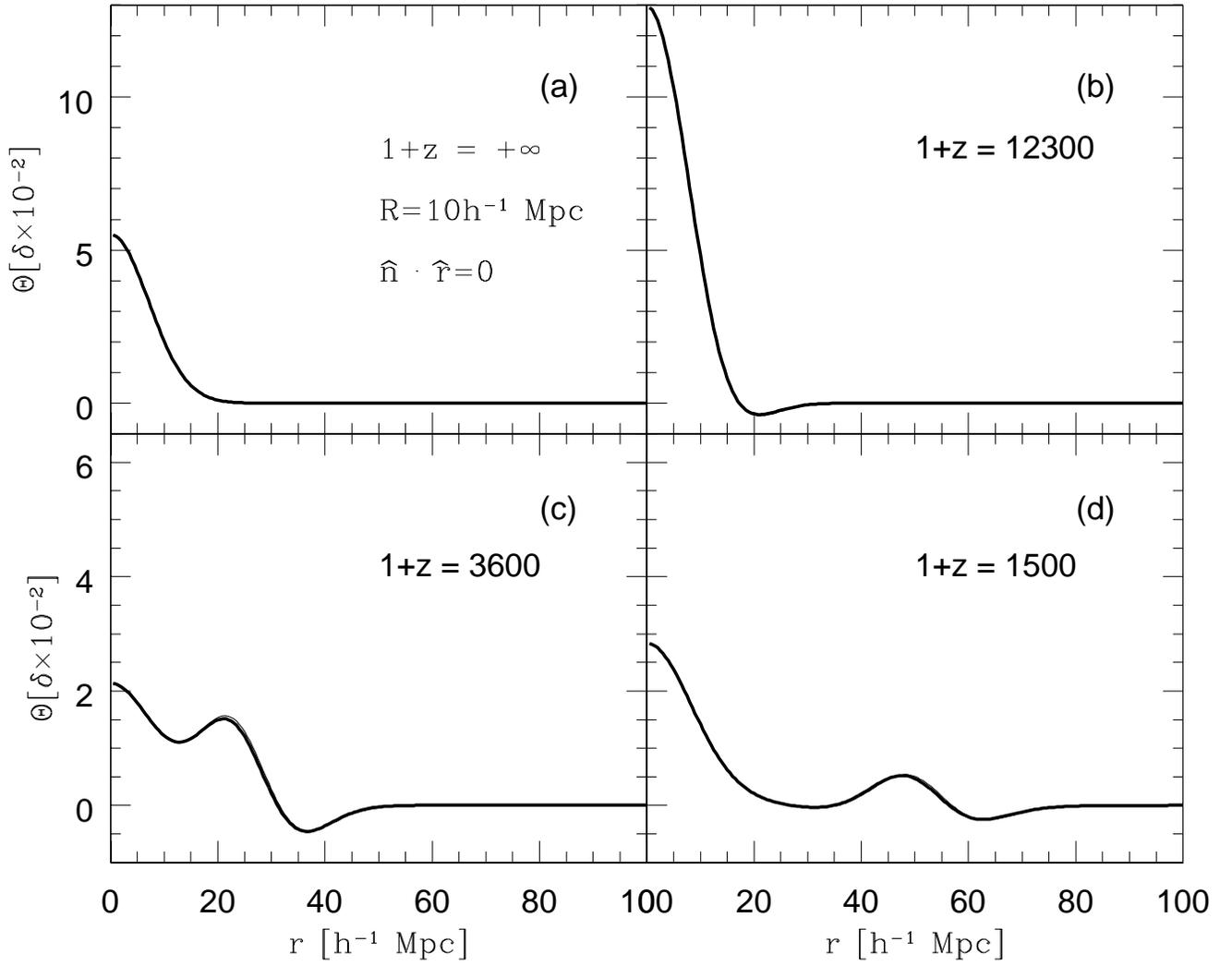}
\caption{CMB temperature perturbation around a 
spherical seed with the indicated size 
as a function of the radial distance from 
the center; the different panels represents the 
perturbation at different times. Note the temperature 
waves arising from the oscillations occurring at the 
horizon crossing.}
\end{figure}

\begin{figure}
\label{p4}
\psfig{figure=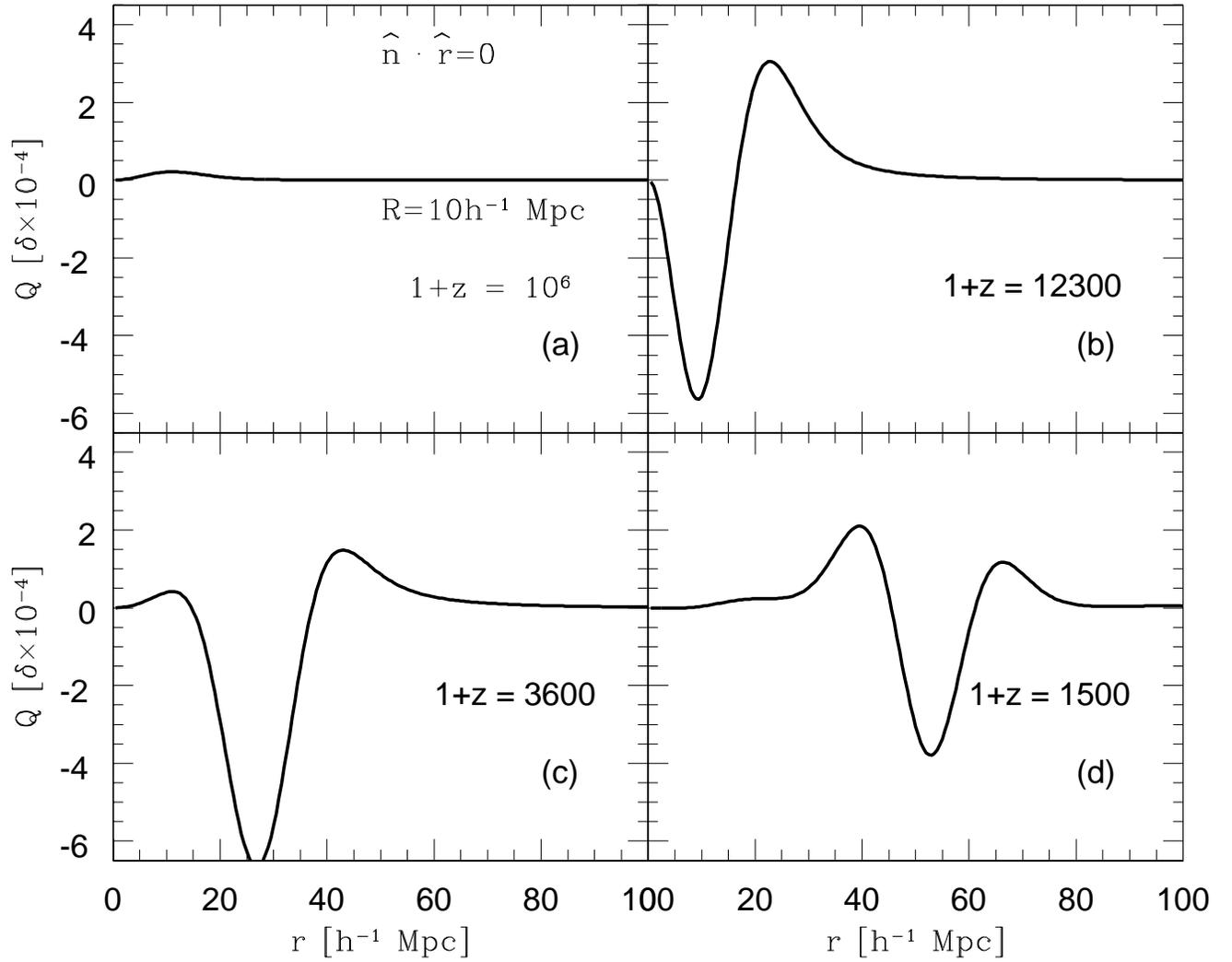}
\caption{The CMB polarization perturbation around a 
spherical seed is plotted as in figure 3. Note the 
external polarization waves at the position of the 
CMB sound horizon at the time considered and the 
absence of central perturbation.}
\end{figure}

\begin{figure}
\label{p5}
\psfig{figure=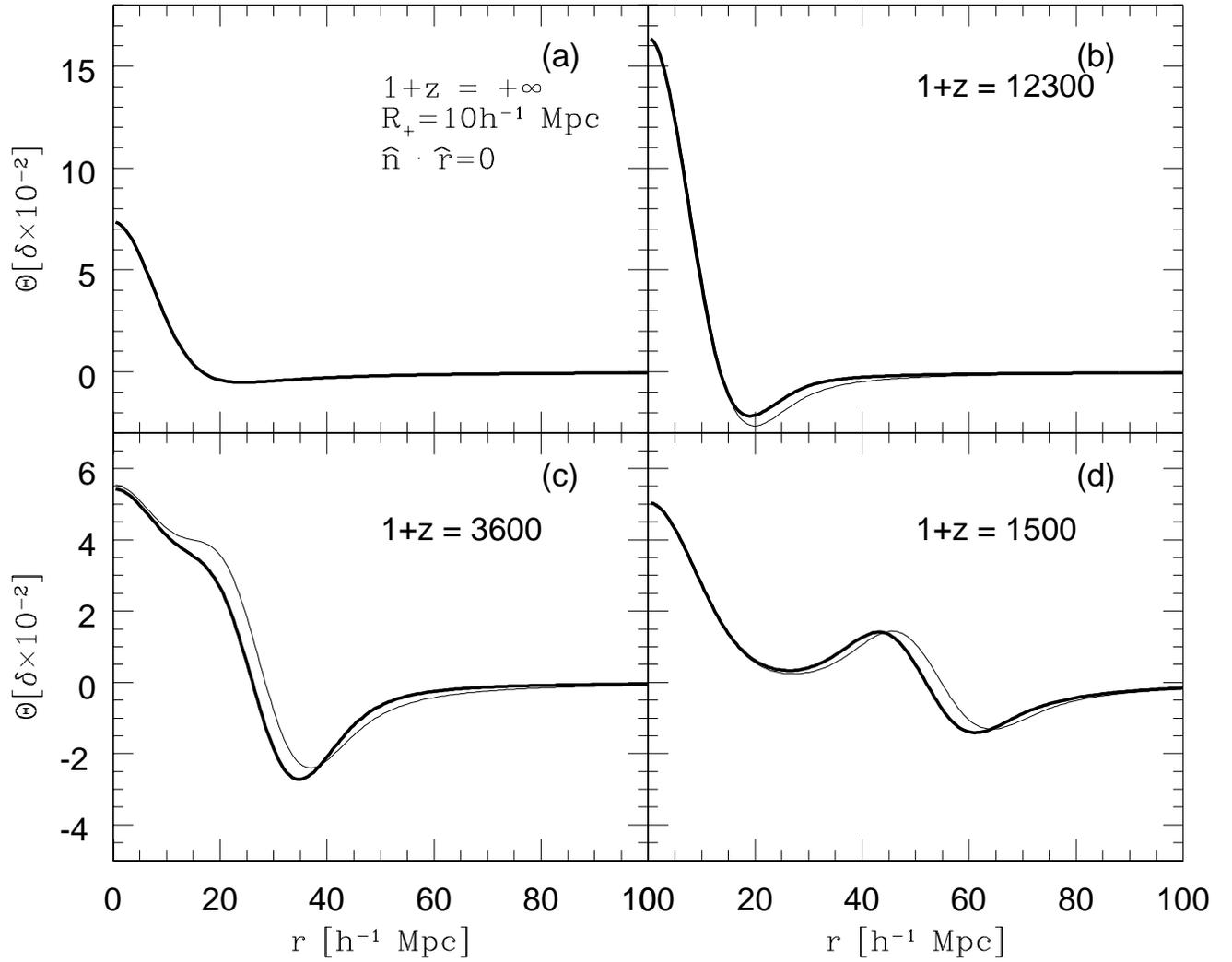}
\caption{CMB temperature perturbation around a 
infinite cylindrical seed with the indicated size on 
the equatorial plane, as a function of the distance 
from the symmetry axis. The different panels represents the 
perturbation at different times. Note, in 
analogy with the spherical case, the temperature 
waves arising from the oscillations occurring at the 
horizon crossing.}
\end{figure}

\begin{figure}
\label{p6}
\psfig{figure=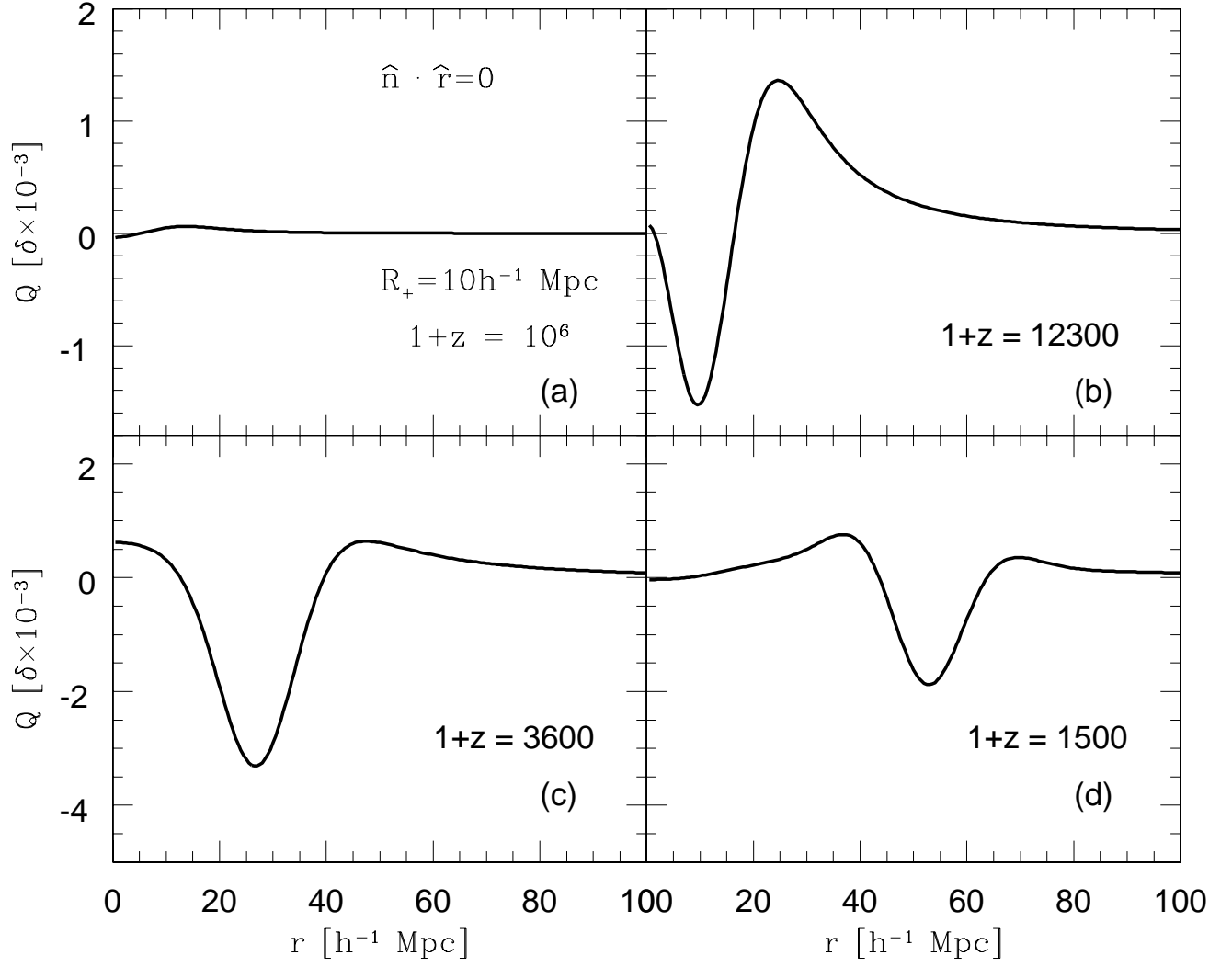}
\caption{The CMB polarization perturbation around a 
infinite cylindrical is plotted as in figure 5. In this 
case, since photons are propagating as indicated, 
a central polarization arises, mostly evident in panel $c$.}
\end{figure}

\begin{figure}
\label{p7}
\psfig{figure=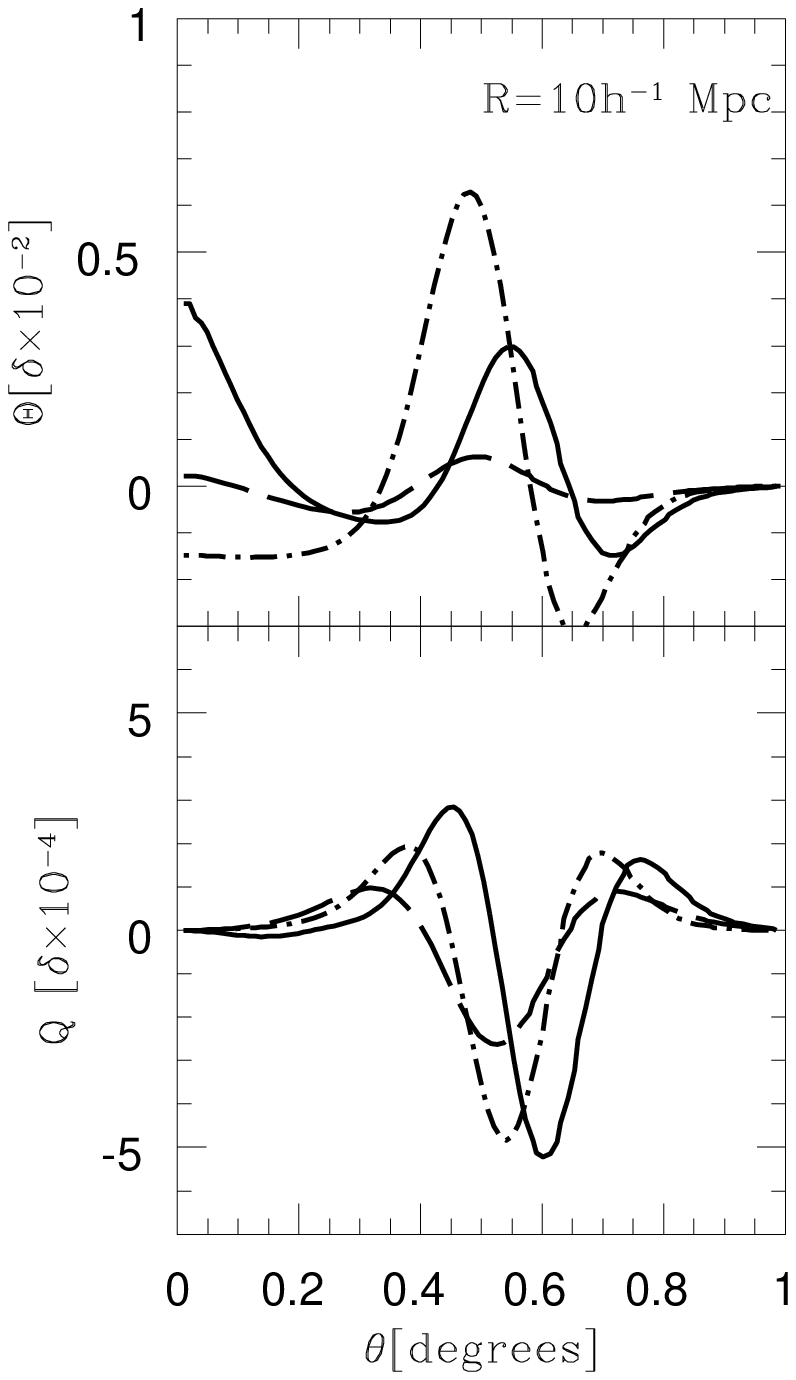}
\caption{CMB temperature (top) and polarization (bottom) 
anisotropy for a spherical seed with the size indicated; 
$\theta$ is the angle from the center. The seed is centered exactly 
on the last scattering surface ($d=0$, solid line), 
just in front of it ($d=30h^{-1}$ Mpc, dashed line), and behind 
($d=-30h^{-1}$ Mpc, dotted dashed line). It physically 
occupies the very central part of the graph, $\theta\le 10'$. 
Note the temperature and 
polarization anisotropy waves at the angular scale corresponding 
to the CMB sound horizon at decoupling.} 
\end{figure}


\begin{thebibliography}{}
\bibitem{B} Baccigalupi C. 1998, Ap.J 496 615, astro-ph/9711095; 
Amendola L., Baccigalupi C. \& Occhionero F. 1998
Ap.J. Lett. 492, L5; Baccigalupi C. 1999, in preparation
\bibitem{BKS} Bardeen J.M. 1980 Phys.Rev.D 22, 1882;
Kodama I. \& Sasaki M. 1984 Progr. of Theor.Phys.Supp. 78, 1 
\bibitem{BW} Born M. \& Wolfe E. {\it Principles of optics}\ , 
Pergamon Press 1980
\bibitem{FOI} Kolb E.W. 1991 Physica Scripta T36, 199
\bibitem{C} Coulson D., 
unpublished Ph.D. Thesis, Princeton University 1994
\bibitem{GR} Gradshteyn I.S. \& Ryzhik I.M.,
{\it TABLES OF INTEGRALS, SERIES, AND PRODUCTS}\ ,
Academic Press 1980
\bibitem{HW} Hu W., Seljak U., White M. \& Zaldarriaga M. 1997, 
Phys.Rev.D 56, 596, astro-ph/9709066; 
Hu W. \& White M. 1997 Phys.Rev.D 56, 596, astro-ph/9702170; 
Hu W. \& Sugiyama N. 1995, Ap.J. 444, 489, astro-ph/9407093 
\bibitem{PRE} Kofman L., Linde A. \& Starobinsky A.
1997, Phys.Rev.D 56, 3258 
\bibitem{BUBBLE} La D. 1991, Phys. Lett. B 265, 232; 
Turner M.S., Weinberg E.J. \& Widrow L.M.
1992, Phys.Rev.D 46, 2384; Occhionero F. \& Amendola L. 1994, 
Phys. Rev. D 50, 4846 
\bibitem{BBS} Kaiser N. \& Stebbins A. 1984, Nat. 310, 391; 
Bouchet F., Bennet D. \& Stebbins A. 1988, Nat. 355, 410 
\bibitem{MFB} Mukhanov V.F. , 
Feldman H.A. \& Brandenberger R.H. 1992 Phys.Rep. 215, 203
\bibitem{CMBPAST} 
Lasenby A.N., Jones A.W. \& Dabrowski Y., 
1998 Phil. Trans. R. Soc. Lond. A. in press, astro-ph/9810196
\bibitem{MB} Ma \& Bertschinger 1995, Ap.J. 455, 7
\bibitem{P} Padmanabhan T.
{\it Structure formation in the universe}\ ,
Cambridge University Press 1993
\bibitem{PLANCK}
Planck Surveyor: 
http://astro.estec.esa.nl/SA-general/Projects/Planck/\ ;\\ 
Microwave Anisotropy Probe: http://map.gsfc.nasa.gov/
\bibitem{TD} Tkachev I., Khlebnikov S., Kofman L. \&
Linde A. 1998,  Phys.Lett. B 440, 262, hep-ph/9805209
\bibitem{CV} White M., Krauss L.M. \& 
Silk J. 1993 Ap.J. 418, 535
\end{thebibliography}
\end{document}